\documentclass[14pt,a4paper]{article}

\usepackage{amsmath}
\usepackage{times}
\usepackage{graphicx}
\usepackage{url}
\usepackage{hyperref}

\begin{document}

\title{Quantifying the `end of history' through a Bayesian Markov-chain approach}

\author
{Florian Klimm$^{1,2\ast}$\\
\\
\normalsize{$^{1}$Department of Computational Molecular Biology, Max Planck Institute for Molecular Genetics,}\\
\normalsize{Ihnestra\ss{}e 63-73, D-14195, Berlin, Germany}\\
\normalsize{$^{2}$Department of Computer Science, Freie Universit\"at Berlin,}\\
\normalsize{Arnimallee 3, D-14195 Berlin, Germany}\\
\\
\normalsize{$^\ast$To whom correspondence should be addressed; E-mail:  klimm@molgen.mpg.de.}
}


\date{}


%
%
%

\maketitle

\begin{abstract}
 Political regimes have been changing throughout human history. After the apparent triumph of liberal democracies at the end of the twentieth century, Francis Fukuyama and others have been arguing that humankind is approaching an `end of history' (EoH) in the form of a universality of liberal democracies. This view has been challenged by recent developments that seem to indicate the rise of defective democracies across the globe. There has been no attempt to quantify the expected EoH with a statistical approach. In this study, we model the transition between political regimes as a Markov process and ---using a Bayesian inference approach--- we estimate the transition probabilities between political regimes from time-series data describing the evolution of political regimes from 1800--2018. We then compute the steady state for this Markov process which represents a mathematical abstraction of the EoH and predicts that approximately $46\,\%$ of countries will be full democracies. Furthermore, we find that, under our model, the fraction of autocracies in the world is expected to increase for the next half-century before it declines. Using random-walk theory, we then estimate survival curves of different types of regimes and estimate characteristic lifetimes of democracies and autocracies of 244 years and 69 years, respectively. Quantifying the expected EoH allows us to challenge common beliefs about the nature of political equilibria. Specifically, we find no statistical evidence that the EoH constitutes a fixed, complete omnipresence of democratic regimes.

\end{abstract}

\section*{Introduction}

Political systems undergo constant changes, which are driven by a variety of internal and external forces~\cite{geddes1999we}. Naturally, this raises the question whether there is an end-point in this development of human societies. Many authors, amongst them Georg Wilhelm Friedrich Hegel~\cite{hegel1837werke}, Karl Marx~\cite{marx2009manifest}, and Karl Popper~\cite{popper2020open}, have been aiming to predict theoretically which kind of political system may constitute this final state of human's sociocultural evolution. Francis Fukuyama popularised the term `end of history' (hereafter EoH), first in a 1989 essay~\cite{fukuyama1989end} and second, in a 1992 book~\cite{fukuyama2006end}, indicating that after the defeat of fascism and communism, the Western liberal democracy may become a universal form of government. He argued that liberal democracies and its accompanying market liberalisation provides a wealth for its citizens that makes transitions from liberal democracies to autocracies unlikely. These ideas are especially challenged in recent years which have seen a rise of so-called hybrid regimes, such as illiberal democracies~\cite{croissant2004introduction,karl1995hybrid}, and a more fine-grained deterioration of democratic norms, jointly referred to as \emph{democratic backsliding}~\cite{mechkova2017much,mounk2018end,wiesner2018stability,runciman2018democracy}, even though a generally agreed definition is lacking~\cite{waldner2018unwelcome}. At least since the 2016 US presidential election, there also has been discussion about the impact of polarisation~\cite{bednar2021polarization}, ethnic antagonism~\cite{bartels2020ethnic,clayton2021elite}, and spread of misinformation~\cite{watts2021measuring} on political decision making, illustrating a rising interest in quantitative studies of the democratic process~\cite{bednar2015resilience,wang2021systems}. Complex-system approaches, in particular, have been used to find hidden structure in political data~\cite{zhang2008community,porter2005network,mucha2010community,lazer2011networks,victor2017oxford}. 

In this manuscript, we use an empirical complex-system approach to predict the EoH and quantify how the EoH would look like under the assumption that the historically observed developments of regimes are representative of the long-term behaviour of their transitions. Specifically, we use a \emph{Markov-chain approach} to model the transition of regimes, which we characterise in an ordinal twenty-one point scale. Markov chains are a ubiquitous tool for statistical data analysis~\cite{behrends2000introduction} and have been, for example, employed to speech recognition~\cite{rabiner1989tutorial} and are also used by Google to rank webpages with their \emph{PageRank} algorithm~\cite{page2001method}. In a political science context, Markov models have been employed to investigate the democratisation process~\cite{gleditsch1997double}, to study, for example whether certain socio-economic factors (e.g., gross domestic product)~\cite{epstein2006democratic} or international influences~\cite{gleditsch2006diffusion} impact the democratisation process. One challenge that is commonly ignored in such transition models is that the available transition data is sparse. We tackle this challenge by using a Bayesian estimator of Markov-chains to infer regime-transition probabilities (For an introduction to Bayesian data analysis methods, see~\cite{gelman1995bayesian}).

The remainder of the manuscript is structured as follows. First, we give a brief, intuitive primer on Markov chains and illustrate them with a simplified model of regime change. Second, we present the results of our analysis based on empirical regime change data. Third, we discuss our findings and limitations. We provide a detailed description of the data and the mathematical methods in the Methods section. The Supplementary Information contains statistical tests and additional results.

\section{A primer on Markov processes with an illustrative model of regime transitions} \label{sec:primer}

\begin{figure}[t]
\centering
\includegraphics[width=.45\linewidth]{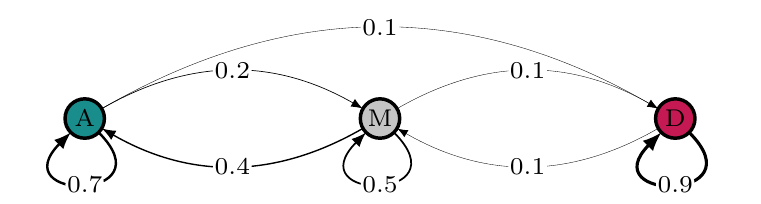}
\includegraphics[width=.225\linewidth]{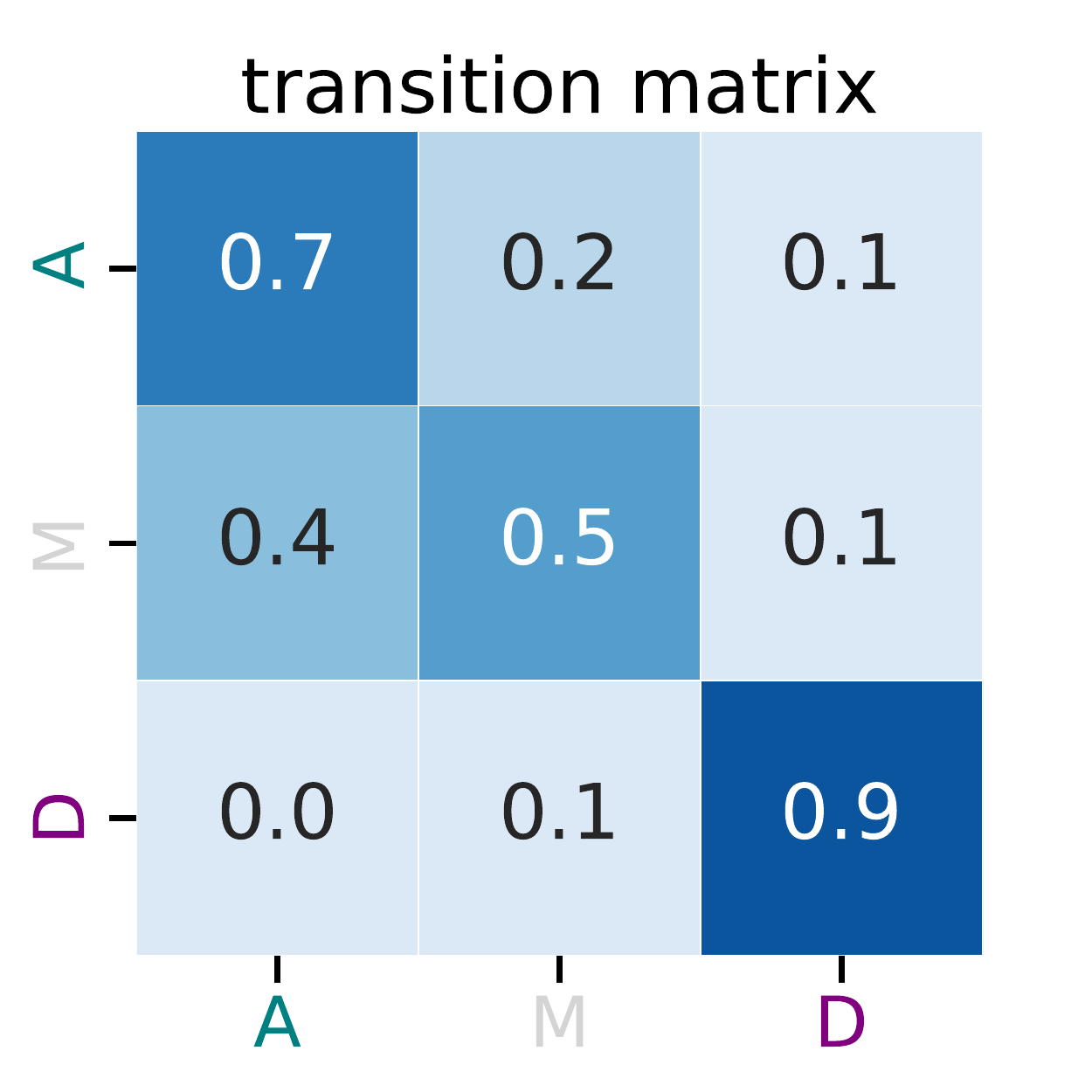}
\includegraphics[width=.225\linewidth]{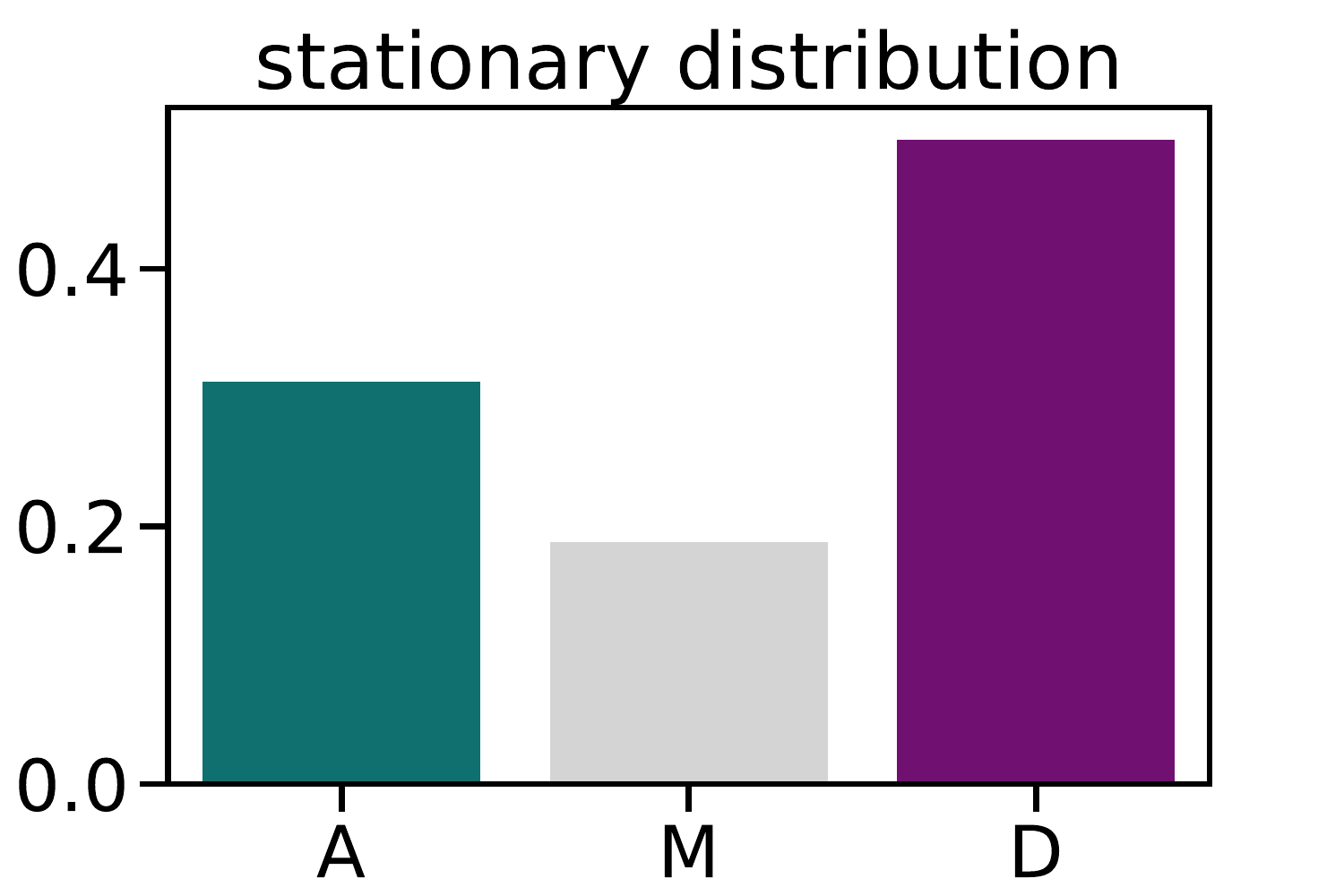}
\caption{We discuss the transition between political regimes as a Markov chain with three states, reprinting autocracies (A), mixed regimes (M), and democracies (D). The transition probabilities between these states are given as arrows. These transition probabilities can also be represented in a transition matrix $\mathbf{P}$. The Markov chain reaches a stationary distribution in which the proportion of the regimes are not changed under the transitions. In this example, the stationary distribution consists of approximately $50\,\%$ democracies, $30\,\%$ autocracies, and $20\,\%$ in mixed regimes.}
\label{fig:markovExample}
\end{figure}

In this section, we provide a brief, intuitive description of Markov processes and how they can be used to model the transition between political regimes. For a more formal discussion, see the Methods section.

Assume the political regime of a country can be in one of three states\footnote{We use the term `state' exclusively to mean `Markov state' and refrain from using it to describe a polity under a system of governance.}: autocratic (A), mixed (M)\footnote{Regimes that mix characteristics of democratic and autocratic regimes are also sometimes referred to as anocracies or semi-democracies.}, or democratic (D). The regime may change over time and we quantify its state yearly. From one year to the next, the regime has a certain probability to stay in the current state or change to one of the two other states. We assume that these transition probabilities depend exclusively on the current state, something called the \emph{Markov property}. 

In Fig.~\ref{fig:markovExample} we give an example of such a Markov process with conveniently (but unrealistically) chosen probabilities of regime changes: The disks represent the three possible states (A, M, and D) and the arrows represent transitions from one to another. In the given example, a mixed regime has a $40\,\%$ chance of becoming an autocracy in the following year, a $50\,\%$ chance of staying a mixed regime, and a $10\,\%$ chance of becoming a democracy. A democracy has a probability of $90\,\%$ of remaining a democracy and a $10\,\%$ probability of becoming a mixed regime. The probability of becoming an autocracy is zero and therefore this transition is not possible. An autocracy has a $70\,\%$ chance of remaining an autocracy, a $20\,\%$ chance of becoming a mixed regime, and a $10\,\%$ chance of becoming a democracy. 

This model is a much simplified description of the actual underling processes which are complex socioeconomical systems that are---most likely---intractable at a global scale. Nevertheless, this abstract description allows us to make certain predictions. Here, we focus on a concept called the \emph{stationary distribution} which represents the distribution of political regimes that is unchanged under the given transition probabilities (for mathematical details on how to compute this see the Methods section). In the example given here, this stationary distribution consists of $5/16 \approx 31\,\%$ autocracies, $3/16 \approx 19\,\%$ mixed regimes, and $8/16 = 50\,\%$ democracies (see bar chart in Fig.~\ref{fig:markovExample}). This stationary distribution, however, does not reflect a situation in which there are no more transitions occurring. Rather, the expected transitions between the different regimes occur at rates that cancel each other out.

\section{Results}

We discuss the transition between political regimes with a Markov model similar to the one discussed in section~\ref{sec:primer}. The model, however, is more complex as it has twenty-one instead of three discrete characterisations of political regimes. These characterisation of political regimes is given by the POLITY2 score~\cite{marshall2019political,plumper2010level}. The POLITY2 score characterises the political regimes of 195 countries on a twenty-one-point scale from $-10$ (least democratic; full autocracy) to $+10$ (most democratic; full democracy) from 1800 to 2018 on a yearly basis. For details on the definition of the data, see the Method section. We separate this data into time series for each country that describe the development of its political regimes.

\begin{figure}[t]{
      \centering 
  \includegraphics[width=0.45\linewidth]{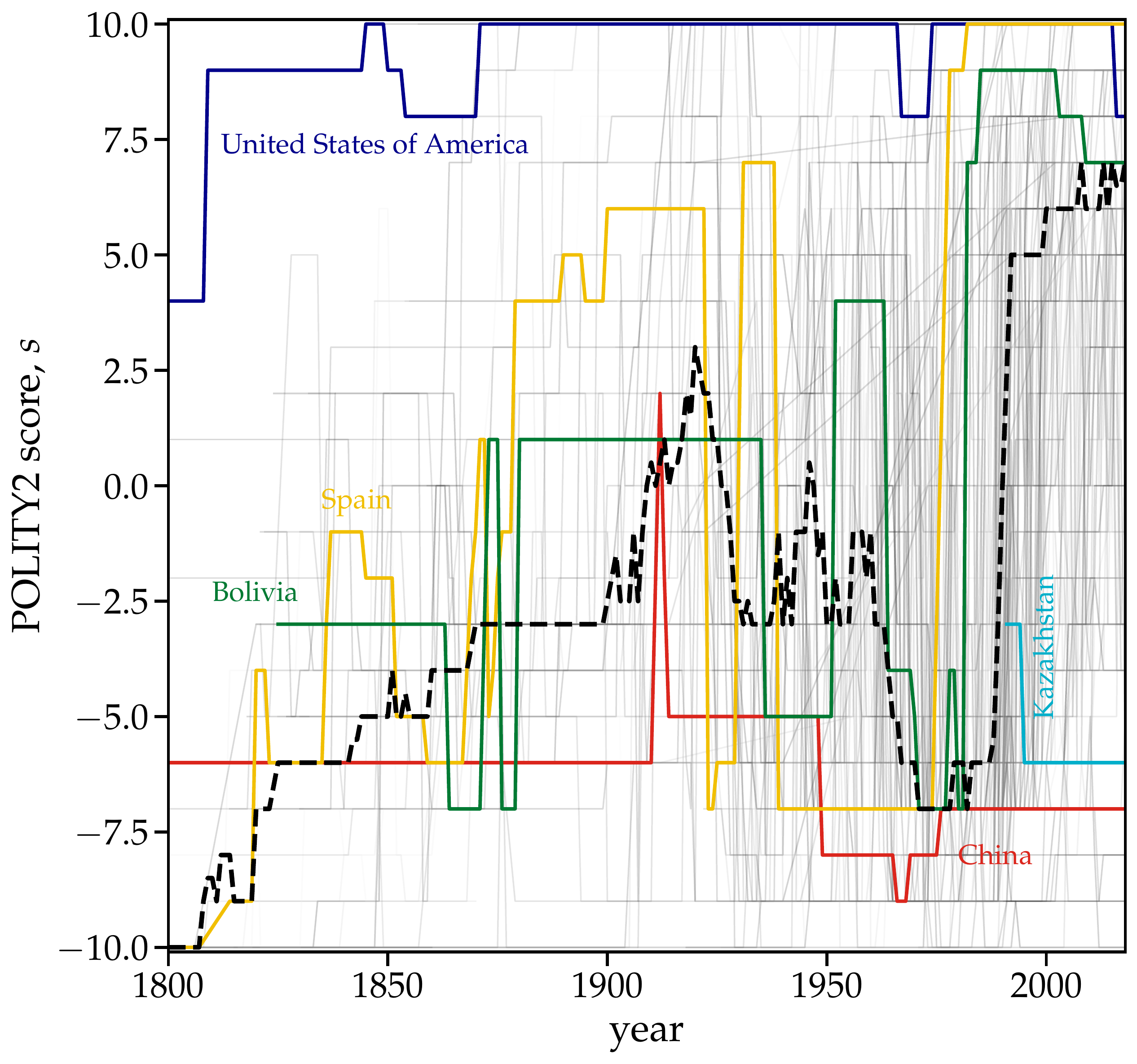}
\caption{The POLITY2 data describes the development of countries' political systems from 1800 to 2018. The POLITY2 score $s$ ranges from $-10$ (least democratic) to $+10$ (most democratic). We show the associated time series of $n=193$ countries and highlight selected countries (United States of America, Bolivia, China, Kazakhstan, and Spain). In addition, we illustrate the temporal development of the median POLITY2 score $\langle s \rangle$ as a dashed line.
   }
   \label{fig:timeSeries}
  }
  
\end{figure}

In Fig.~\ref{fig:timeSeries}a, we show the obtained times series describing $n=193$ countries. The time series differ in their length with a mean length of $\langle L \rangle \approx 89$ years. We highlight the time series for five selected countries, which differ drastically in their POLITY2 score. The United States of America have a relatively high POLITY2 score $s$ but show a recent decline. Spain's POLITY2 score $s$ has been increasing and decreasing over time but reached recently its maximum $s=+10$. Kazakhstan, a former Soviet republic, has a much lower POLITY2 score $s=-5$.

Modelling the time series of the POLITY2 score as a Markov model assumes that a countries transition probability depends exclusively on the current POLITY2 score. While this is a common assumption in many analyses of regime transitions (e.g., \cite{epstein2005higher,epstein2006democratic,ratner2009reaping,albertus2012coercive,gassebner2013extreme}), it is a drastic simplification. To test whether this is a reasonable assumption, we use a statistical procedure for the estimation of the order of a Markov chain from time-series data~\cite{petrovic2022learning}. We compute the Akaike information criterion (AIC) and the Bayesian Information Criterion (BIC) for Markov-chain models of order $K=1,\dots, 7$. The information criteria identify which of different models describe data best (for details see Supplementary Information~7). Our analysis yields that a Markov chain of first order (i.e., a memoryless Markov model) describes the data best. In the following, we will therefore assume that our data follows a memoryless Markov model.

\subsection{Countries have a predominantly constant POLITY2 score but tend to become more democratic over time}

 We use a Bayesian mean posterior approach to estimate the transition probabilities between the twenty-one states from these time series under the Markov assumption. As the data of regime transitions is sparse, we use a Bayesian approach to update our believe in the transition probabilities (see Methods section). The methodological advantages of using a Bayesian approach are two-fold. First, it allows us to estimate underlying transition probabilities, even though some regime transition occur rarely. Second, it allows us to obtain a unique stationary distribution, which we will discuss in more detail in subsection~\ref{subsec:EoH}.

\begin{figure}[t]{
      \centering 
  \includegraphics[width=0.5\linewidth]{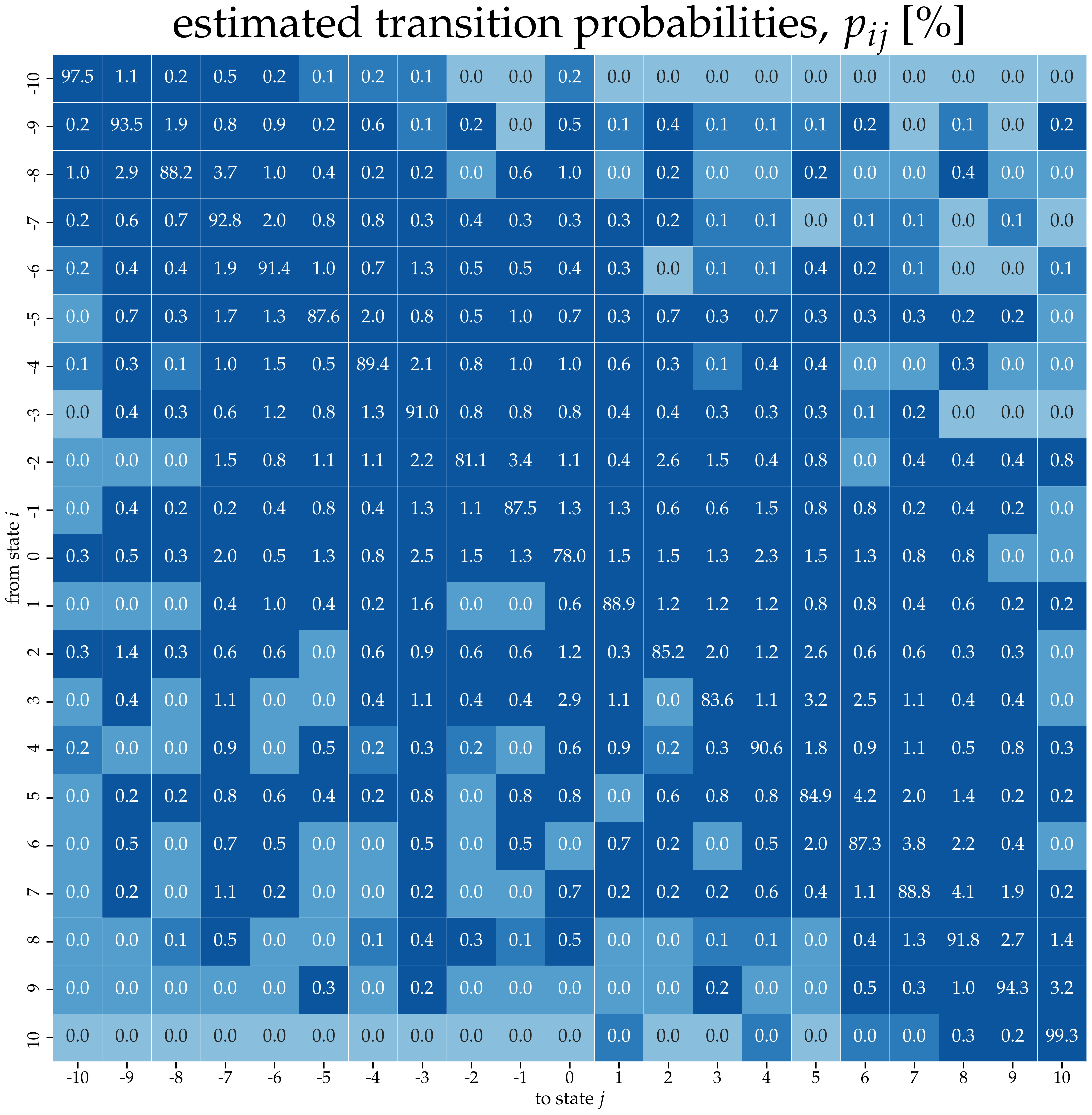}
\caption{Countries' regimes are predominantly constant as the estimated transition matrix $\mathbf{P}$ has high probabilities along the diagonal. Entries above the diagonal represent the probability of regime transitions that increase the POLITY2 score (i.e., more democratic) and entries below the diagonal represent regime transitions that decrease the POLITY2 score (i.e., less democratic). The transition probabilities $p_{ij}$ are given in percent and colour-coded from low (bright) to high (dark).
   }
   \label{fig:transitionMatrix}
  }
  
\end{figure}

We show the transition matrix $\mathbf{P}$ in Fig.~\ref{fig:transitionMatrix}. The matrix is dominated by its diagonal elements, which indicates that regimes have a high probability of staying at the same POLITY2 score. A country with score $-10$ (full autocracy), for example, has a probability of $97.5\,\%$ to stay at this score in the following year. The other diagonal elements are similarly high,  with the lowest being $p_{00} \approx 78\,\%$. The probabilities of regime changes are represented by the off-diagonal elements in $\mathbf{P}$ with the elements above the diagonal representing an increasing POLITY2 score (i.e., becoming more democratic) and the elements below the diagonal representing a decreasing POLITY2 score (i.e., becoming less democratic). We observe that both transition directions are possible, but transitions that increase the POLITY2 score tend to be more likely. On average, across all regimes, the POLITY2 score is expected to increase by approximately $\langle p_{ij} \rangle \approx 0.4$ per year, which is in accord with earlier results on data up to the end of the twentieth century~\cite{gleditsch1997double}. While transitions that change the POLITY2 score strongly (e.g., from $-9$ to $+10$ with probability $0.2\,\%$) are possible, smaller transitions (e.g., from $-9$ to $-8$ with probability $1.9\,\%$) tend to be more likely. The most likely transition that increases the POLITY2 score is from $+5$ to $+6$ and has a probability of $4.2\,\%$. The most likely transitions that decreases the POLITY2 score are from $+3$ to $0$ and from $-7$ to $-8$, both with a probability of $2.9\,\%$. An example of a fairly likely off-diagonal transition that changes the POLITY2 score strongly is the transition from $+2$ to $-9$ with a probability of $1.4\,\%$, a transition that occurred, for example, in 1898 in Guatemala, when Manuel Estrada Cabrera established a dictatorship~\cite{rendon1988manuel}.

\clearpage

\subsection{Full democracies and full autocracies are more stable than mixed regimes}

\begin{figure}[t]{
      \centering 
  \includegraphics[width=0.5\linewidth]{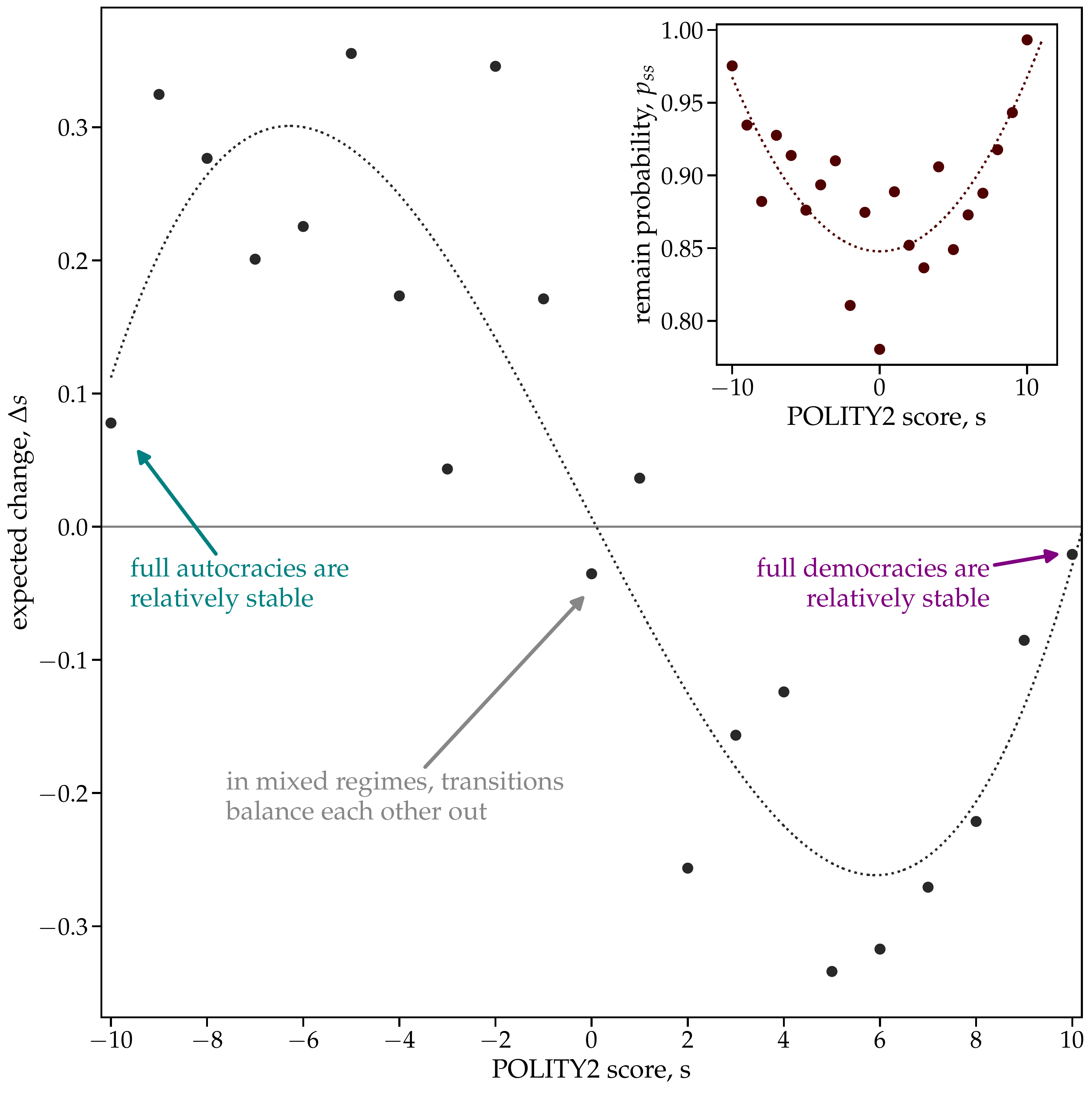}
\caption{The expected change $\Delta s$ of the POLITY2 score follows a cubic polynomial with countries being full autocracies ($s=-10$), mixed regimes ($s=0$), or full democracies ($s=+10$) having expected changes $\Delta s$ close to zero. Regimes that are autocratic ($s<0$) are expected to increase their POLITY2 score, whereas regimes that are democratic ($s>0$) are expected to decrease their POLITY2 score. We show the cubical trend line as a least-square fit ($R^2\approx 0.84$). \emph{(Inlay:)} The remain probabilities $p_{ss}$ are smallest for mixed regimes ($s=0$) and highest for extreme regimes ($s=-10$ and $s=10$), resulting approximately in a quadratic polynomial (see trend line, $R^2\approx 0.63$).
\label{fig:deltaS}
  }
  }
\end{figure}

For each state $i$, we compute the expected change of the POLITY2 score as $\Delta s(i)=\sum_{j=1}^N p_{ij}(i-j)$~(see Fig.~\ref{fig:deltaS}). The expected change $\Delta s$ of the POLICY2 scores is close to zero for full autocracies ($s=-10$), mixed regimes ($s=0$), and full democracies ($s=+10$). The empirically obtained $\Delta s(i)$ therefore resembles a cubic polynomial, which we highlight through a least-square fit. This cubic curve can be understood from the remain probabilities $p_{ss}$, which are the diagonal elements of the transition matrix $\mathbf{P}$ (see inlay in Fig.~\ref{fig:deltaS}). We find that $p_{ss}$ follows approximately a parabola with the extreme regimes (i.e., full autocracies and full democracies) having remain probabilities $p_{ss}$ close to $100\,\%$, resulting in small expected changes $\Delta s \approx 0$. Mixed regimes have the smallest remain probability $p_{00}\approx 78\,\%$, yet the expected change $\Delta s \approx 0$ as transitions that increase and decrease the POLITY2 score $s$ tend to balance each other out. A simple mathematical model describing the resilience of extreme regimes is

\begin{align*}
	p_{ij} &=  \begin{cases}
		p_{\mathrm{remain}}(i)&\mathrm{if}\ i=j\,,\\
		\frac{1-p_{\mathrm{remain}}(i)}{2L} &\mathrm{if}\ i\neq j\,,
	\end{cases}
\end{align*}
with $p_{\mathrm{remain}} (i) = p_0+p_2\left(i/L\right)^2$, for which we compute the expected change to
\begin{align*}
	 \Delta s(i)= \frac{N}{2L}\left( p_{\mathrm{remain}}(i)-1\right)i = \frac{N}{2L}\left( p_0+p_2\left(i/L\right)^2-1\right)i\,,
\end{align*}
resulting in a cubic polynomial, as observed in the inferred transition matrix $\mathbf{P}$, indicating that the particular form of $\Delta s(i)$ is indeed driven by the large remain probabilities $p_{ss}$ of extreme regimes.


\subsection{At the EoH, a plurality but no majority of countries are predicted to be democracies} \label{subsec:EoH}

Under the Markov process described by the transition matrix $\mathbf{P}$, countries may constantly change their POLITY2 score: Some regime changes make countries more democratic and other changes make countries less democratic. As we estimated that autocracies have a higher probability of becoming democracies than democracies becoming autocracies we expect the amount of democracies to increase over time. When there are much more democracies than autocracies, however, a dynamic equilibrium is established in which the rate of democracies dying is the same as the rate of autocracies becoming democracies. Necessarily, there must be a distribution of regimes that is unchanged under the transition matrix $\mathbf{P}$ and this is called the \emph{stationary distribution} (see Methods section). Once this stationary distribution is reached, it does not change anymore and therefore represents a mathematical abstraction of the EoH. The stationary distribution is unique for irreducible and aperiodic Markov chains. As we used a Bayesian estimator, necessarily all transition probabilities $p_{ij}>0$, making the obtained Markov chain irreducible and aperiodic, yielding a unique EoH. We note that methodologically different but conceptually similar ideas, in the form of the random-surfer model, have been employed by Google in the PageRank algorithm~\cite{page2001method}.

\begin{figure}[b!]{
      \centering 
  \includegraphics[width=0.5\linewidth]{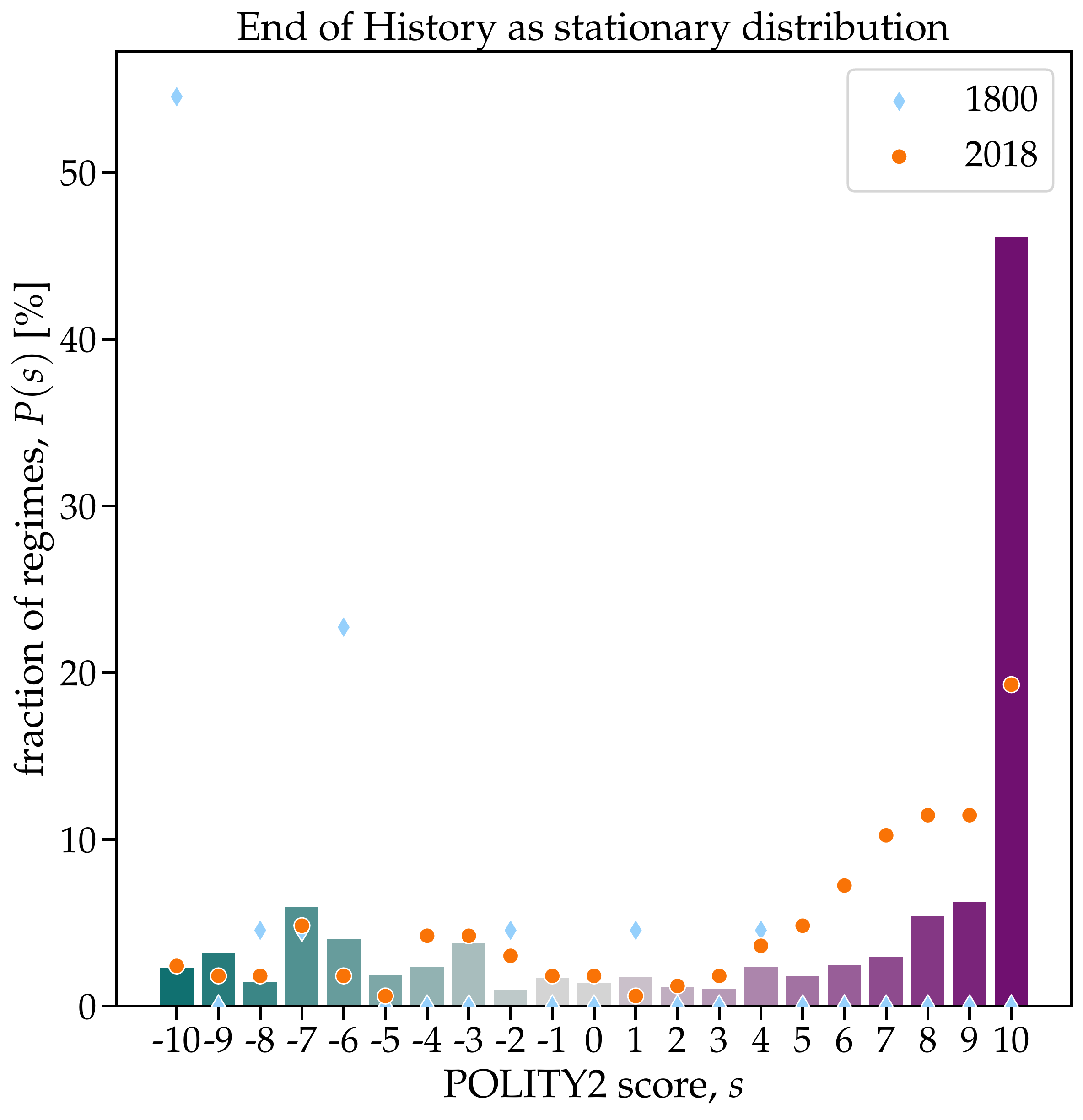}
\caption{At the EoH, approximately $50\,\%$ of countries are expected to be full democracies (i.e., POLITY2 score $s=10$). 
We show the fraction of regimes that has has each POLITY2 score for 1800 (blue diamonds) and 2018 (orange discs). While the fraction $P(s=10)$ of full democracies is in 2018 considerably lower than at the EoH, the fraction of democracies with intermediate POLITY2 score $3\leq s\leq 9$ is lower at the EoH.
\label{fig:steadyState}
  }
  }
\end{figure}

We show the stationary distribution $\vec{\pi}$ of the estimated transition matrix $\mathbf{P}$ in Fig.~\ref{fig:steadyState}. We see that $46\,\%$ of all regimes are predicted to be full democracies (i.e., $s=+10$). The fraction of full autocracies ($s=-10$) is with $2\,\%$ much smaller. The fraction of countries in any state of autocracy (i.e., $ s \leq 0$) is, however, with $28\,\%$ sizeable. The remaining $26\,\%$ of countries are in various states of partial democracies (i.e., $0<s<10$).

We can compare this predicted distribution at the EoH with the observed distributions in the years 1800 and 2018 (shown as blue diamonds and orange disks, respectively). We do observe that the fraction of full autocracies did indeed shrink over these two centuries, while the number of full democracies did rise. At the EoH the fraction of full democracies is expected to be considerably higher than in 2018, while the fraction of partial democracies is expected to be smaller.

To investigate the robustness of our results under different choices in the data analysis, we explore the EoH in slightly different variations in the Supplementary Information. All of these support our finding that a plurality of countries is expected to be a full democracy. In Supplementary Information 2, we compute the EoH for the POLITY score, which is an older version of the POLITY2 score. In Supplementary Information 3, we compute the EoH for different binning of the POLITY2 score. In Supplementary Information 4, we investigate the \emph{Varieties of Democracy} score. In Supplementary Information 5, we study the EoH for data covering different temporal subsets of the regime-transiton data and find that the EoH is largely temporally invariant. In Supplementary Information~10, we study whether the EoH is robust under counterfactual perturbations of the time-series data (in particular no collapse of the USSR and a stronger democratisation during the Arab Spring) and find that such perturbations alter the results negligibly.

\clearpage

\subsection{The amount of autocracies is predicted to reach a maximum in 2063}

\begin{figure}[b!]{
      \centering 
  \includegraphics[width=0.5\linewidth]{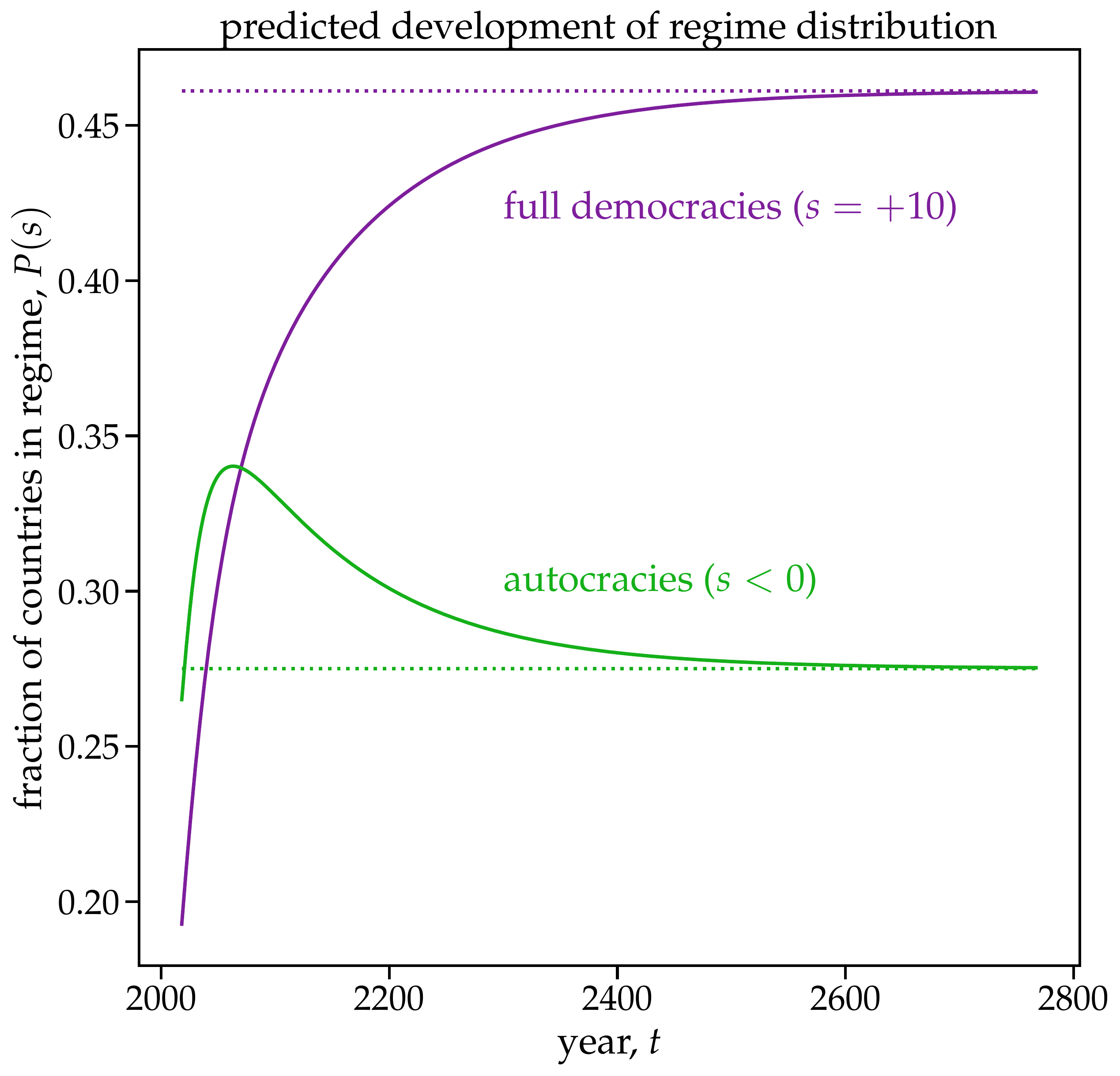}
\caption{While the predicted fraction of full democracies increases over time and approaches the steady state, the fraction of autocracies increases for $43$ years and decreases then to reach the steady state.
We show the fraction of full democracies (i.e., $s=+10$) as purple line and the fraction of autocracies (i.e., $s<0$) as a green line. Both approach their respective fractions (shown as dashed lines) at the EoH without ever reaching it but get within $1\,\%$ in approximately 400 years. 
\label{fig:development}
  }
  }
\end{figure}

Using the transition matrix $\mathbf{P}$, describing the Markov model, we can investigate the expected development of the distribution of regimes extrapolating from the last available data for the year 2018. We expect that the EoH in the form of the unique stationary distribution $\vec{\pi}$ of regimes is approached for $t \to \infty$. In Fig.~\ref{fig:development}, we show the expected temporal development of the distribution of POLITY2 scores over the next 800 years. Specifically, we show the fraction of full democracies (i.e., $s=+10$) and the fraction of autocracies (i.e., $s<0$). We see that both approach their respective steady-state distributions at the EoH without ever exactly reaching it. While the fraction of full democracies is continuously increasing over time, the fraction of autocracies is increasing until 2063, when it reaches a maximum of $34\,\%$, and then shrinking until it reaches the steady state of $28\,\%$. In the year 2070, the fraction of full democracies is predicted to be larger than the fraction of autocracies for the first time.

\clearpage

\subsection{The expected lifetime of a democracy is threefold that of an autocracy}

As established earlier, the expected steady state at the EoH is not characterised by an absence of regime change rather there is an equilibrium between democracies becoming autocracies and autocracies becoming democracies. We can aim to characterise these fluctuations at the EoH. Specifically, we can numerically estimate the expected lifetime of democracies and the expected time until an autocracy becomes a full democracy as \emph{hitting times} for random walks (see Methods section).

For this, we simulate the trajectories of $r=10^5$ political regimes that start as full democracies (i.e., in state $s=+10$). We simulate the transitions as Markov process given by the inferred transition matrix $\mathbf{P}$. We compute the time $t$ until these countries become an autocratic regime (i.e., reach a state $s<0$). We estimate the median lifetime $t_{1/2}^{\mathrm{autocracy}}$ of a full democracy of $244$ years. In the same say, we may estimate the median lifetime $t_{1/2}^{\mathrm{autocracy}} \approx 69$ years of a full autocracy by starting $r=10^5$ political regimes in the state $s=-10$ and compute the time until a democracy (i.e., $s>0$) is reached. The lifetimes for both type of regimes have relatively large standard deviations with $62$ years and $310$ years, for autocracies and democracies, respectively.

\begin{figure}[b!]{
      \centering 
  \includegraphics[width=0.45\linewidth]{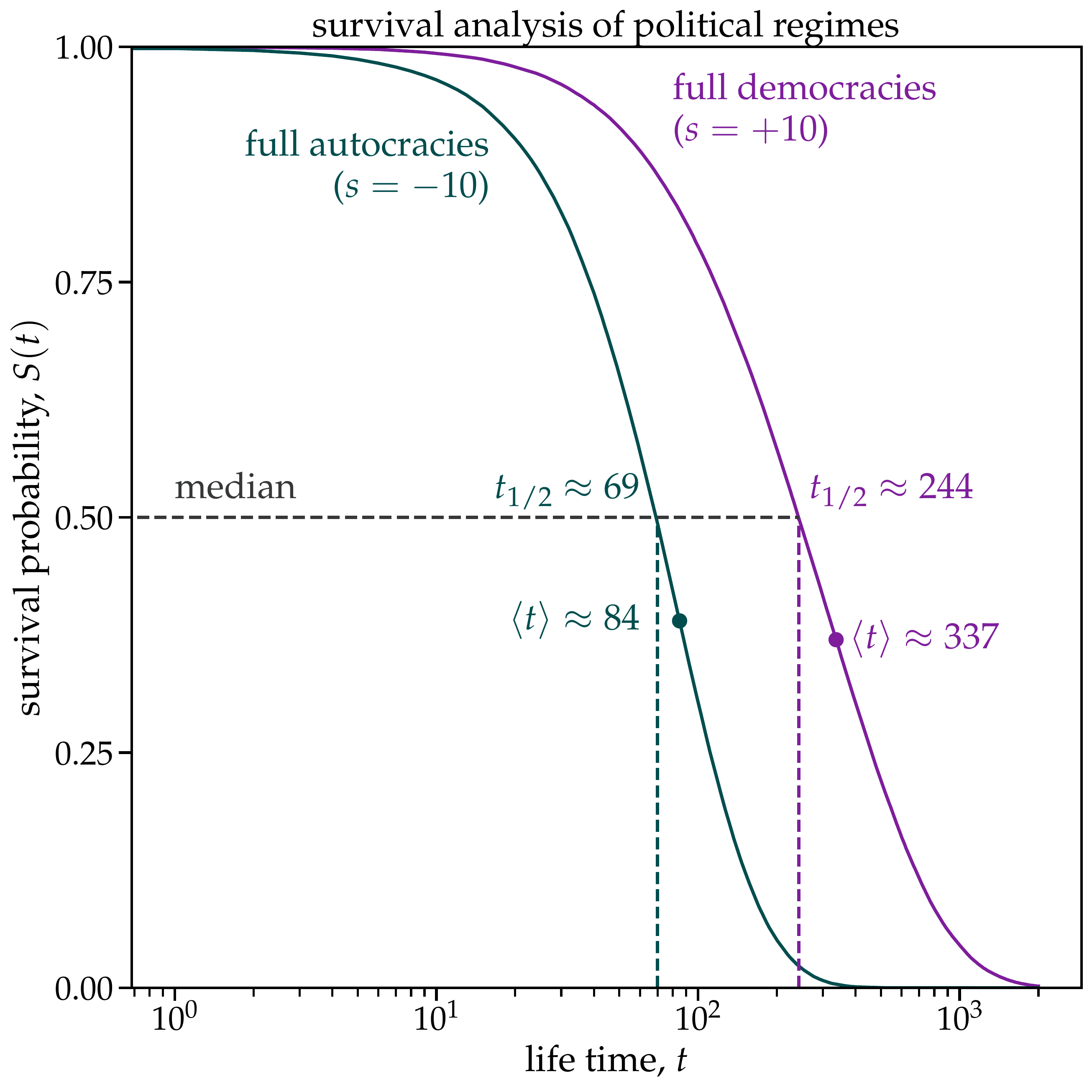}
\caption{The survival probability $S(t)$ for full autocracies drops quicker than for full democracies. We estimate the median lifetime of both types of regimes to $t_{1/2}^{\mathrm{autocracy}} \approx 69$ years and $t_{1/2}^{\mathrm{democracy}} \approx 244$ years, respectively. The survival probabilities $S(t)$ of both regimes were estimated with a Kaplan--Meier estimator applied to $10^5$ time series. We may also determine the mean lifetimes of regimes from the transition matrix directly by computing hitting times of the Markov chain explicitly. We obtain mean lifetimes of full autocracies and full democracies of $\langle t \rangle^{\mathrm{autocracy}} \approx 84$ and $\langle t \rangle^{\mathrm{democracy}} \approx 337$, respectively.
  }
  }
  \label{fig:survivalCurve}
\end{figure}

More generally, we may estimate the survival function $S(t)$ of political regimes, which gives the probability that a regime survives past time $t$. In Fig.~\ref{fig:survivalCurve}, we show the Kaplan--Meier~\cite{kaplan1958nonparametric} estimator of the survival function $S(t)$ for these simulated regimes (see Methods section).  We find that the full autocracies' survival probability $S(t)$ drops considerably faster than the one of full democracies. The survival probabilities fall below $1\,\%$ after 287 and 1450 years for full autocracies and full democracies, respectively.

The Markov-chain model allows us to compute the mean lifetimes of regimes from the transition matrix directly. For this, we compute hitting times $\tau_{i,A}$ by solving the associated system of linear equations (see Methods section).
The mean time until a full democracy becomes an autocracy for the first time is then, for example, $\tau_{+10,s<0}$ and we obtain $\tau_{+10,s<0}=\langle t \rangle^{\mathrm{democracy}} \approx 337$. Analogously, we compute the mean lifetime of a full autocracy to $\tau_{-10,s>0}=\langle t \rangle^{\mathrm{autocracy}} \approx 84$, indicating that the mean lifetime of a full democracy is four-fold the lifetime of a full autocracy. We verify these analytical expressions by comparing them with the numerical estimates $\hat{\tau}$ and find that they differ by less than a year for both, autocracies and democracies. For both type of regimes, the mean lifetime and the median lifetime are of similar magnitude, yet $\langle t \rangle > t_{1/2}$, which indicates that a small number of the simulated regimes have much larger lifetimes than the majority of regimes. In the Supplementary Information 7, we show that an empirical cumulative distribution function yields virtually indistinguishable results as the Kaplan--Meier estimator.


\section{Conclusions}

In this manuscript, we used a Markov-chain approach to estimate the transition probabilities between political regimes from time-series data covering more than two centuries. We found that the most extreme regimes (i.e., full autocracies and full democracies) have the highest probabilities to persist. Using the estimated transition probabilities allowed us to quantify the distribution of political regimes at the EoH as a stationary distribution of the Markov chain. We find that approximately $49\,\%$ of countries are predicted to become full democracies with a median lifetime of about $244 \pm 310$ years. Autocracies make up $26\,\%$ of regimes and have a median lifetime of $69 \pm 62$ years. 

Analysing the predicted temporal development from 2018 until the EoH, we find a steady increase of full democracies and a steady decline in the number of partial democracies. Surprisingly, we also detect an increase of autocracies for the next 50 years, which are followed by a decline thereof. This development is mainly driven by a current large number of partial democracies which are more likely to become autocracies than full democracies over the short term, even though they might become full democracies in long term. This indicates that the currently observed democratic backsliding might be a harbinger of further incline of partial democracies becoming autocracies, even if the EoH is characterised with a larger number of full democracies.

In our approach, we treated all countries the same and did not consider country-specific factors that could influence democratisation. Other studies have demonstrated that there is evidence for the `modernisation hypothesis' (i.e., that countries that do economically well tend to undergo liberal democratic transitions)~\cite{epstein2006democratic,diamond1989democracy,przeworski1997modernization,inglehart2005christian}, influence of state legacies on democratisation~\cite{hariri2012autocratic,sinding2019power,hariri2015contribution}, and a correlation between scientific production and democracy~\cite{guigo2019correlation}. The modernisation hypothesis in particular, could mean that the transition probabilities are not fixed but rather are a function of a countries prosperity, which in turn makes the EoH dependent on the economical development. Such approaches that take into account additional socio-economical factors can be fruitful and lead to a deeper understanding of the drivers of democratisation. Nevertheless, we refrain from doing so in this study, because this would require us to also predict the development of these external factors until the EoH, which would lead to a much more complex model with many parameters, making it likely intractable. Similarly, it has been shown that the spread of regime change can be influenced by international treaties (e.g., defensive alliance)~\cite{cranmer2020contagion}, an effect we refrain from including in our model. To some extent, we believe that the strength of our model is not its (potential) accuracy of prediction but rather its simplicity. While the prediction of quantities, such as the time until the EoH, should not be mistaken as an exact forecast, they nevertheless give indications about the order of magnitudes of the expected outcomes. To touch on the international variability, we compute EoHs for specific countries or regions and find a large variability, with some regions being largely democratic and others being dominated by autocracies (see Supplementary Information~1).

Independent of the exact predicted values, our findings support a view of the EoH in which there is not a single, omnipresent type of political regime but rather a broad range of regimes that span all values across the POLITY2 scale. In particular, we find no statistical evidence for a universality of full liberal democracies as suggested by Francis Fukuyama, although we predict them to represent a plurality of political regimes. Yet, in accordance with Francis Fukuyama, we find that full democracies are the most stable type of regime.

In our model, we treat each country's regime as an entity independent of others. It is known, however, that there are considerable influences between countries that might lead to drastic changes in many countries in a short amount of time a so-called `Democratic Domino Theory'. This occurred, for example, during the decline of European democracies in the 1930s or the fall of the Soviet Union at the end of the Cold War and there is some empirical evidence for such spatial-temporal effects~\cite{leeson2009democratic,o1998diffusion}, raising the question to what extent such synchronised dynamics might perturb the EoH. The influence of such synchronisation events might also lead to more complex dynamical processes which are not fully described by our stationary Markov chain, making the investigation of more advanced models a fruitful endeavour. 

Our data covers predominantly the last two centuries, which ---despite some considerable setbacks--- have been a success story of liberal democracies, most likely driven by a drastically rising average income. The analysis of this time-period most likely leads to an overestimation of transition probabilities that increase the POLITY2 score. With reliable time series data that cover more years, we would be able to extend the analysis presented here. Furthermore, we might update the estimation of the transition matrix with more recent data, once available, indicating a need for further empirical studies of regime transitions and their long-term progression.

\section*{Materials \& Methods}

\subsection*{Markov Chain}

%
%

We analyse the time series of countries' POLITY2 scores as a finite sequence of random variables $\vec{X}=$ $(X_1$, $X_2$, $\dots$ ,$X_n)$. Let $\mathcal{X}$ be the state space of random variables, i.e., the set of values that each variable can take. In our case this is the POLITY2 score $s$ and therefore $X_i \in \mathcal{X} = [-10,+10]$ for all $i$. The size $N=21$ of the state space is number of possible POLITY2 scores. We treat these time series as discrete-time \emph{Markov Chains}.

A finite discrete-time Markov chain is a sequence of random variables $(X_1$, $X_2$, $\dots$ ,$X_n)$ that fulfil the Markov property.

The Markov property is also called `memoryless property' because it requires that the transition probability only depends on the current state. Specifically, the probability $P( X_{t+1} = y | X_t = x_t,X_{t-1} = x_{t-1},\dots,X_1= x_1)$ of being in state $x$ conditioned on the whole history of states only depends on the last state, i.e., $P( X_{t+1} = y | X_t,\dots,X_1)  = P( X_{t+1} = x | X_t = x_t) = p_{x_ty}$.


We can describe a Markov chain through a \emph{transition matrix} $\mathbf{P}$ that describes the probability of a transition from one state to another. Specifically, $\mathbf{P}$ is a non-symmetric $N \times N$ matrix, where $N$ is the number of states (in our case the 21 possible POLITY2 scores) and entry $p_{ij} \in [0,1]$ indicates the probability of transition from state $i$ to state $j$. The rows of $\mathbf{P}$ each sum to $1$ (i.e., $\sum_{j=1}^N p_{ij} =1$) because the probabilities are normalised.

\subsection*{Bayesian estimation of Markov Process from time-series data}

In our application, we do not have the transition matrix $\mathbf{P}$ given \emph{a priori}. Rather, we want to estimate it from observed time series. Let $\vec{X}^{(1)}$, $\vec{X}^{(2)}$, $\dots$ be a set of sequences of random variables, each representing the time series of political regime characterisation for one country. 

Let $n_{ij}$ be the number of times that we observe a transition from state $i$ to state $j$ in the time series. In a frequentist approach the maximum-likelihood estimation~\cite{grimmett2020probability} of the transition probability from state $i$ to state $j$ is then

\begin{align}
	\hat{p}_{ij} = \frac{n_{ij}}{n_{i+}} = \frac{n_{ij}}{\sum_{j=1}^{N} n_{ij}}\,,
\label{eq:estimatorFreq}
\end{align}
where $n_{i+}$ is the total number of observed transitions starting in state $i$. This estimator is a so-called consistent estimator. 

For our purposes, it can be beneficial to use a Bayesian approach to estimate the transition matrix. One reason is that just because we have never observed a certain transition, we do not expect the probability of this transition to occur in future is zero. Rather we would like to assign it a small but finite probability. From a Bayesian approach this follows naturally by combining prior beliefs with the observed data. 

We assign each row of the transition matrix a Dirichlet prior with equal weighting of the states such that $\alpha_{ij}=1/N=1/21$ for all $(i,j)$, such that each transition has the same probability. We update the prior belief with the observed data and obtain a posterior mean estimate 

\begin{align}
	\hat{p}_{ij}^{\mathrm{(Bayes)}} = \frac{n_{ij} +\alpha_{ij}}{\sum_{j=1}^{N} (n_{ij}+\alpha_{ij})}\,.
\label{eq:estimatorBay}
\end{align}
 
 The posterior mean estimate $\hat{p}_{ij}^{\mathrm{(Bayes)}}$ converges to frequentist maximum-likelihood estimation $\hat{p}_{ij}$ for large amounts of data. Only when data is scarce, the Bayesian estimate adds corrections to the frequentist approach~\cite{strelioff2007inferring}. For example, a transition $i \to j$ that is never observed such that $n_{ij}=0$, still receives a small finite probability $p_{ij}>0$. We show the transition matrix estimated with a frequentist approach in Supplementary Information~6. We note that the used estimator resembles the rule of succession, as introduced by Pierre-Simon Laplace~\cite{marquis1840essai}.

\subsection*{Stationary distribution}
Let $\vec{x}(t) = (x_1, x_2, \dots, x_N)$ be a state vector that indicates the distribution among the $N$ states at time $t$. 
We can compute the evolution of a Markov chain by multiplying a state vector $\vec{x}(t)$ with the transition matrix $\mathbf{P}$ such that the state vector at time $t+1$ is

\begin{align}
	\vec{x}(t+1) = \vec{x}(t)\mathbf{P}\,.
\end{align}

A \emph{stationary distribution} $\vec{\pi}$ is a state vector that does not change under the multiplication with the transition matrix $\mathbf{P}$ such that 
\begin{align}
	\vec{\pi} = \vec{\pi}\mathbf{P}\,.
\end{align}

That is, once a Markov chain reached a stationary distribution, it stays there. Every irreducible and aperiodic Markov chain has a unique stationary distribution (for details see~\cite{behrends2000introduction}). Both conditions are fulfilled in our case. In particular, the irreducibility is necessarily (i.e., independent of the actual time series data) fulfilled because we use the Bayesian estimator (Eq.~\ref{eq:estimatorBay}). The frequentist estimator (Eq.~\ref{eq:estimatorFreq}) could lead to a reducible Markov as some transitions are never observed.

We can compute the stationary distribution $\vec{\pi}$  as the eigenvector of the transition matrix $\mathbf{P}$ with eigenvalue $1$. There are many numerical algorithms able to calculate the eigenvectors and we use the QR algorithm as used by standard {\sc Python} linear algebra library.

\subsection*{Hitting time}

The \emph{hitting time}, also called first passage time, is the mean number of steps $\tau _{ij}$ it takes a random walker starting at node $i$ to visit a target set of nodes $A$ for the first time~\cite{maier2019modular}. For a Markov chain, we can compute this explicitly by solving the system of linear equations
\begin{align*}
	\tau_{i,A} = \begin{cases}
		1 + \sum_{j \in S} p_{ij} \tau_{j,A}&\mathrm{if}\ i \notin A\\
		0 &\mathrm{if}\ i \in A\,.
	\end{cases}
\end{align*} 

We also may estimate it numerically by simulating a large number of random walks starting from node $i$ and choosing random transitions in accordance with the transition matrix $\mathbf{P}$. We stop the random walk when we reach the target set of nodes $A$ for the first time. The number of steps taken until this happens is called the length of the walk. Assume, we have simulated $r$ random walks starting a $i$ and ending at $j$ and $L(w)$ indicates the length of walk $w$. We then estimate the hitting time to

\begin{align}
	\hat{\tau}_{ij} = \frac{\sum_{w=1}^r L(w)}{r}\,.
\end{align} 

\subsection*{Kaplan--Meier estimator}

We estimate the survival function $S(t)$ of political regimes from simulated regime-transition time series. In particular, we simulate $r=10^5$ political regimes and obtain their life times as hitting times. We use the non-parametric Kaplan--Meier estimator~\cite{kaplan1958nonparametric}

\begin{align}
	\hat{S}(t) = \prod_{i: t_i\leq t}^{\infty} \left(1-\frac{d_i}{n_i}\right)\,,
\end{align}
where $d_i$ are the number of political regimes dying at time $t_i$ and $n_i$ are the number of regimes that have survived until time $t_i$.

\subsection*{Data}

We extract time series data from `POLITY5: Political Regime Characteristics and Transitions, 1800-2018'~\cite{marshall2019political}, which is available under \url{https://www.systemicpeace.org/polityproject.html}. Specifically, we use the Revised Combined Polity Score (POLITY2) score, which is a composite indicator. It is computed by subtracting the AUTOC score from the DEMOC score. The AUTOC score is an eleven-point scale (0--10), which combines measurements of the competitiveness of political participation, the regulation of participation, the openness and competitiveness of executive recruitment, and constraints on the chief executive. The DEMOC score is an eleven-point scale (0--10), which combines measurements data concerning the `Competitiveness of Executive Recruitment', the `Openness of Executive Recruitment', `Constraint on Chief Executive', and the `Competitiveness of Political Participation'. Accordingly, the resulting unified polity scale (POLITY2) ranges from $-10$ (fully autocratic) to $+10$ (fully democratic) and has twenty-one different values $s\in [-10,+10]$.

The data covers 219 years and 195 countries. We treat each country's trajectory as a separate time series. A time series is an ordered vector $\vec{S} = (s_1, s_2, \dots, s_{t'})$ in which $s_i$ indicates the score at time $t$. The length $t'$ of a time series is the number of elements in the vector. As some countries cease to exist or emerge during this period, the time series vary in their length.



\clearpage 

\subsection*{Data \& Code Availability}

All raw data is available under \url{https://www.systemicpeace.org/polityproject.html}. We make {\sc Python} code to reproduce all of our results available under \newline 
\url{https://github.com/floklimm/democracyMarkovChain}. 

\subsection*{Funding}
FK is supported as an Add-on Fellow for Interdisciplinary Life Science by the Joachim Herz Stiftung, by the EPSRC (Centre for Mathematics of Precision Healthcare; EP/N014529/1), and by the Max-Planck Society.

\subsection*{Acknowledgments}
We thank Nick S. Jones, Benjamin F. Maier, Tammo Rukat, and Martin Vingron for fruitful discussions. We thank Luka Petrovi\'{c} and Ingo Scholtes for advice regarding the Markov order of paths in graphs. We are grateful to the comments by two anonymous reviewers, whose recommendations greatly improved the manuscript.


\pagebreak

\bibliography{democracyState}

\bibliographystyle{unsrt}

\end{document}


\maketitle

\section{Regional differences of the `end of history'}

In the main manuscript, we infer the Markov chain transition matrix $P$ from time series from a total of $n=193$ countries. Using this transition matrix $P$, we then compute the EoH as the stationary distribution $\vec{\pi}$ associated with this transition matrix. This approach has the implicit assumption that all countries can be described by the same transition matrix $P$.

In this section, we compute EoHs for different subsets of countries. First, we identify the EoHs associated with countries from specific geographic regions. Second, we compute the EoHs associated with specific countries.

\subsection{Region-specific `End of history'}

We can associated each country with geographic regions and we choose the continents Asia, America, Africa, Europe, and Oceania. While most countries are associated with a single region, some countries span multiple regions (e.g., Russia being in Asia and Europe). 

\begin{figure}[t]
\centering
\includegraphics[width=.80\linewidth]{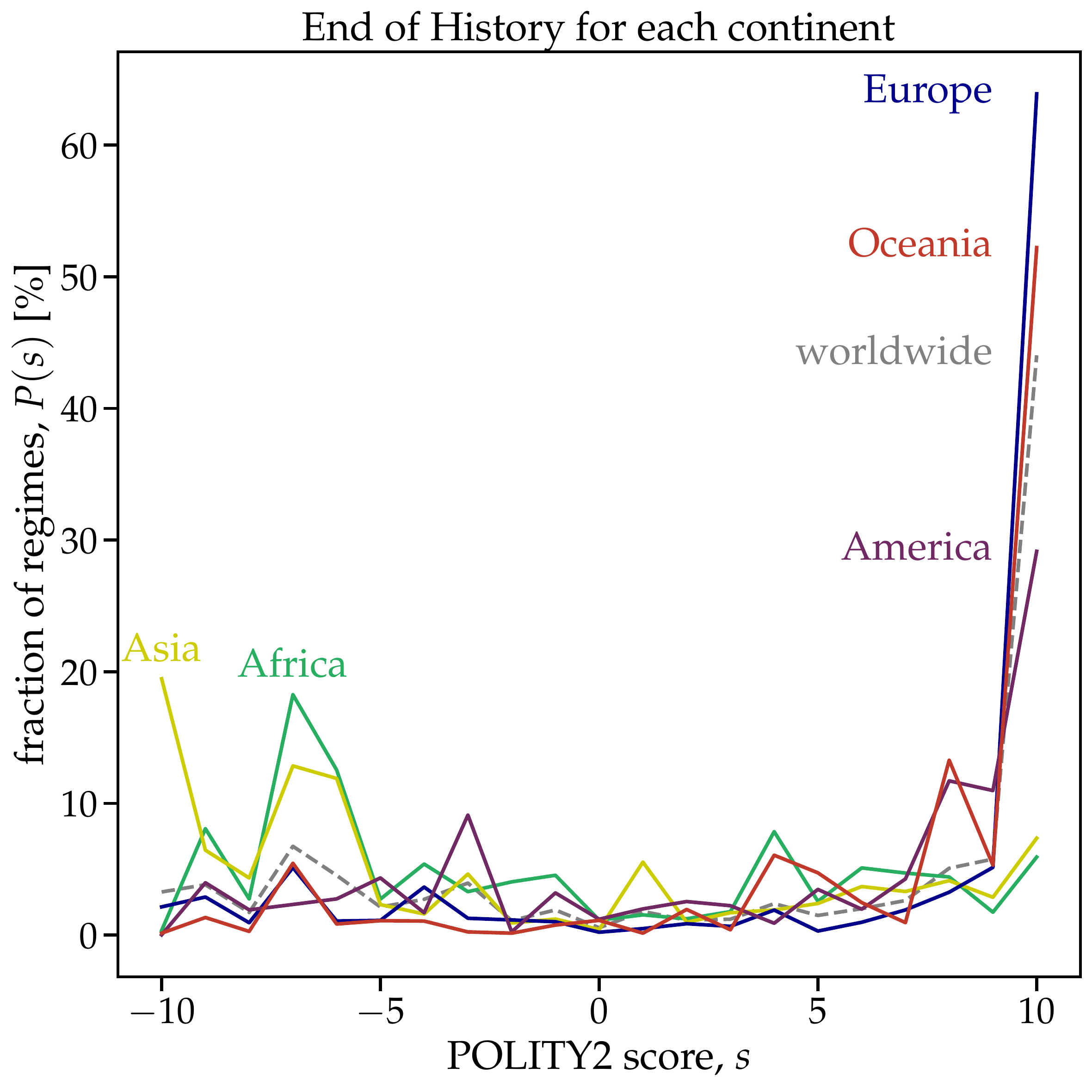}
\caption{
At the EoH, approximately $50\,\%$ of countries are expected to be full democracies (i.e., POLITY2 score $s=10$). 
We show the fraction of regimes that has has each POLITY2 score for 1800 (blue diamonds) and 2018 (orange discs). While the fraction $P(s=10)$ of full democracies is in 2018 considerably lower than at the EoH, the fraction of democracies with intermediate POLITY2 score $3\leq s\leq 9$ is lower at the EoH.
}
\label{fig:continents}
\end{figure}

We use this information to first compute a transition matrix $P$ for each region and second compute the EoH associated to each region. In Fig.~\ref{fig:continents}, we show EoH for each region and the EoH for the aggregated data. We find strong regional influences on the EoH. In particular, America, Oceania, and Europe have the highest fraction of full democracies ($s=+10$), whereas Asia and Africa have the highest fraction of autocracies ($s=-10$ and $s=-7$, respectively). This indicates that there are spatial-proximity effects which yield a region-specific EoH. This finding can be explained with two hypothesis, both of which have some evidence in the democratic transition literature. First, state legacies such as colonisation have influence on democratisation processes~\cite{hariri2012autocratic}. Second, spatial interdependence leads to spatio-temporal clustering of democracy and autocracy~\cite{elkink2013spatial}. 

\subsection{Country-specific `End of history'}

We may also compute country-specific transition matrices in our framework. For this, we consider only a single time series. This approach has the shortcoming that for countries with small amount of data (i.e., short time series), the obtained transition matrix with our Bayesian approach is dominated by the prior. This will lead to a EoH that is almost uniform. 

\begin{figure}[t]
\centering
\includegraphics[width=.6\linewidth]{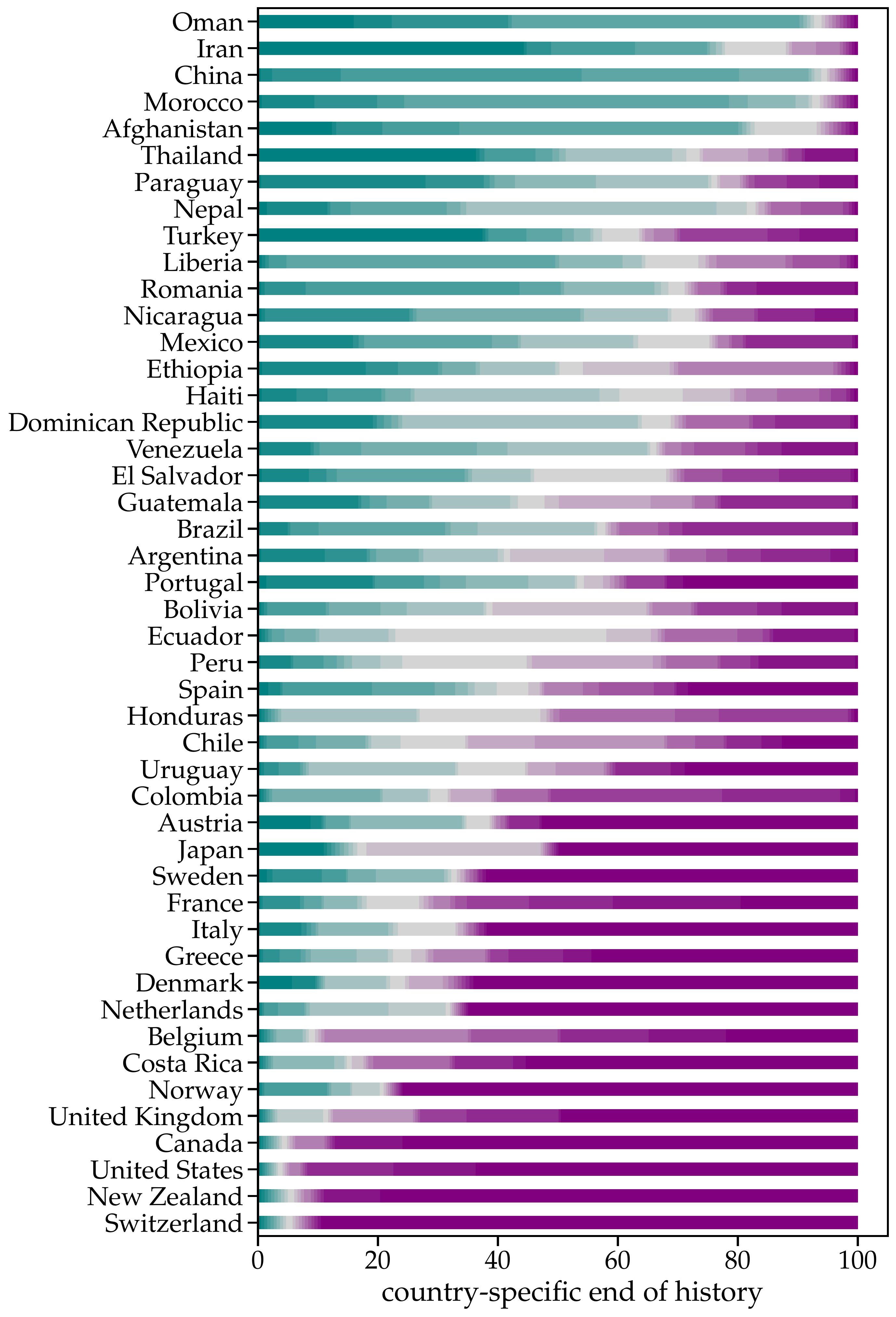}
\caption{Countries vary strongly in their country-specific EoH. We find that some countries` EoH is mainly autocratic and others are mainly democratic. Countries are sorted by the mean POLITY2 score at their EoH from $s=-10$ (top) to $s=+10$ (bottom). We only show countries whose time series covers at least 150 years.}
\label{fig:countrySpecific}
\end{figure}

Thus, we employ this approach only to countries for which we have a time series that cover at least $150$ years. In Fig.~\ref{fig:countrySpecific}, we show the country-specific EoH for these countries. We find that the countries span a wide range of possible EoHs. Countries at the top of the figure have EoHs that are largely autocratic (e.g., Oman, Iran, China, Morocco, Afghanistan). Countries towards the bottom of the figure have EoHs that are largely democratic (e.g., Switzerland, New Zealand, United States, Canada).

\clearpage

\section{`End of history' for the POLITY score}

In the main manuscript, we investigate the EoH from the POLITY2 data. The POLITY2 data is a revised version of the POLITY score. In this section, we investigate to what extent our results hold for the POLITY data.

The POLITY2 score is a revised version of the POLITY data in which some missing data (e.g., for interregnum periods) is imputed with scores. While this approach yields more complete time series, it is also a potential source of error~\cite{plumper2010level}. Both, POLITY and POLITY2, range from $-10$ to $+10$ and thus span 21 different states.

\begin{figure}[h]
\centering
\includegraphics[width=.44\linewidth]{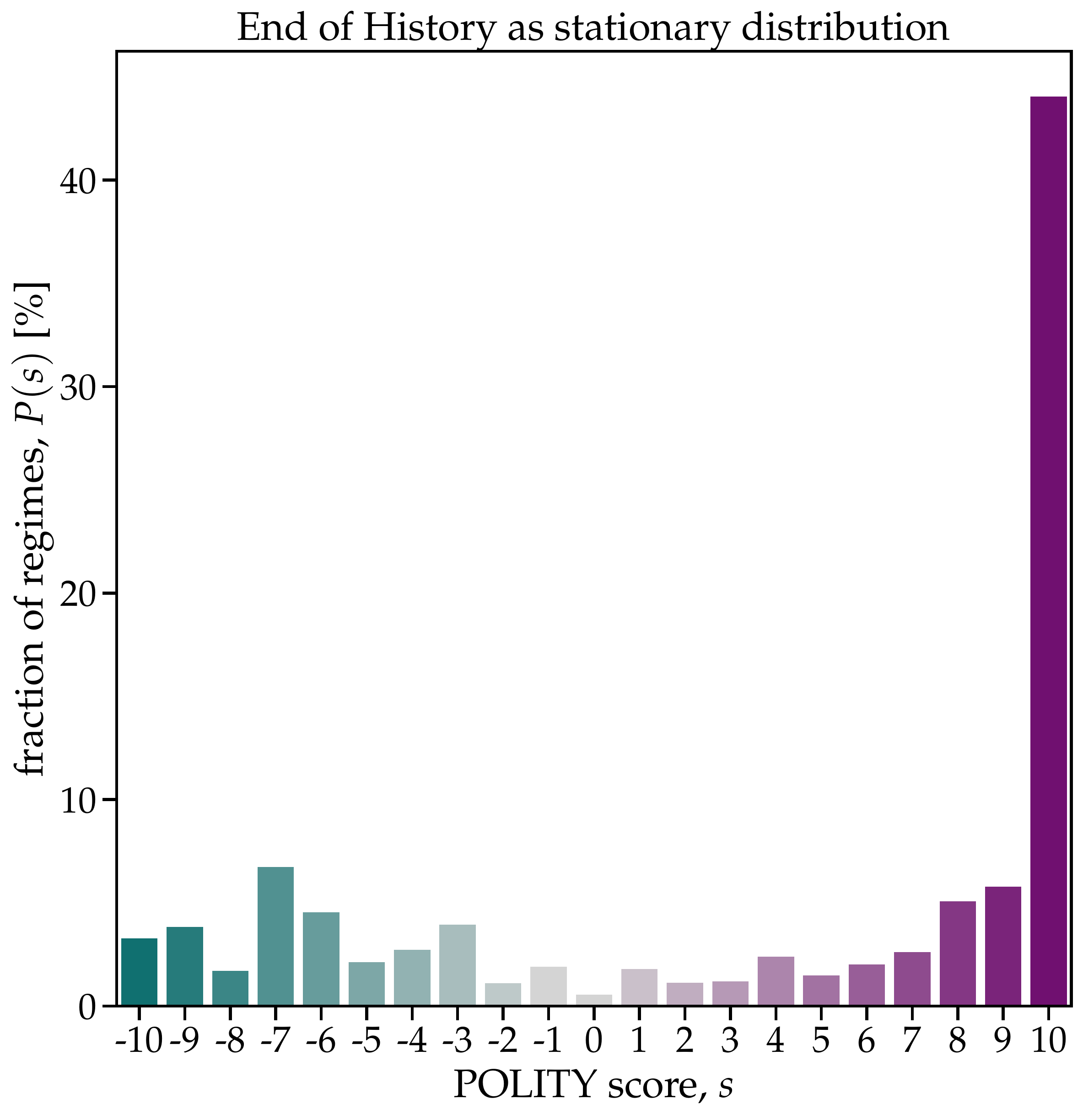}
\includegraphics[width=.49\linewidth]{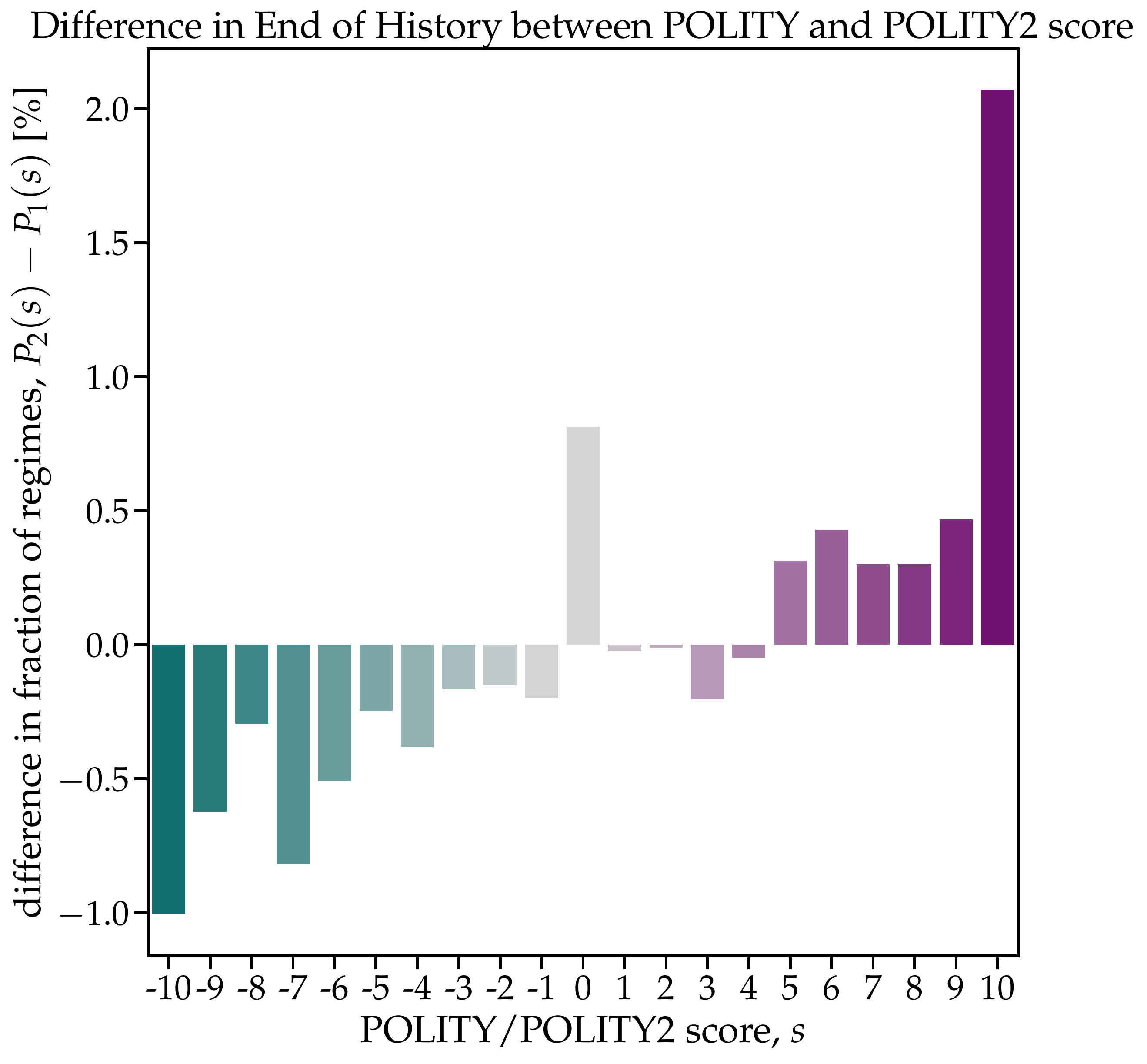}
\caption{The EoH for POLITY2 and POLITY score are very similar. (Left) We show the EoH from the POLITY time series. (Right) We compute the difference between the EoH as estimated from the POLITY score with the EoH as estimated for the POLITY2 score. While there are differences between both, the difference for each state is less than $2\,\%. $}
\label{fig:polity}
\end{figure}

In Fig.~\ref{fig:polity}a, we show the EoH for time series derived from the $POLITY$ data. We find that our main results, a plurality but no majority of countries being full democracies, holds. In Fig.~\ref{fig:polity}b, we explore the difference in the EoH between both data. We identify that the fraction of democracies tend to be slightly larger for the POLITY2 score data.

\clearpage

\section{`End of history' for different binnings of the POLITY2 score}

In the main manuscript, we infer the Markov chain transition matrix $P$ and compute the EoH for the POLITY2 which consists of 21 different states, ranging from $-10$ (most autocratic) to $+10$ (most democratic). In this section, we study to what extent our findings are robust under a  coarse-graining of the data.

\DeclarePairedDelimiter{\nint}\lfloor\rceil

\DeclarePairedDelimiter{\abs}\lvert\rvert

We study two different coarse-graining variants, both of which aggregate multiple states into one state. First, we study a binning into three states $s_3=\{-1,0,+1\}$. We compute this from the original data as $s_3 = \nint[]{s/7}$, where $\nint[]{}$ indicates rounding to the closest integer. Second, we study a binning into seven states $s_7=\{-3,-2,-1,0,+1,+2,+3\}$. We compute this from the original data as $s_7 = \nint[]{s/3}$. In both cases, the smallest value indicates the most autocratic and the largest value indicates the most democratic state.

\begin{figure}[b!]
\centering
\includegraphics[width=.49\linewidth]{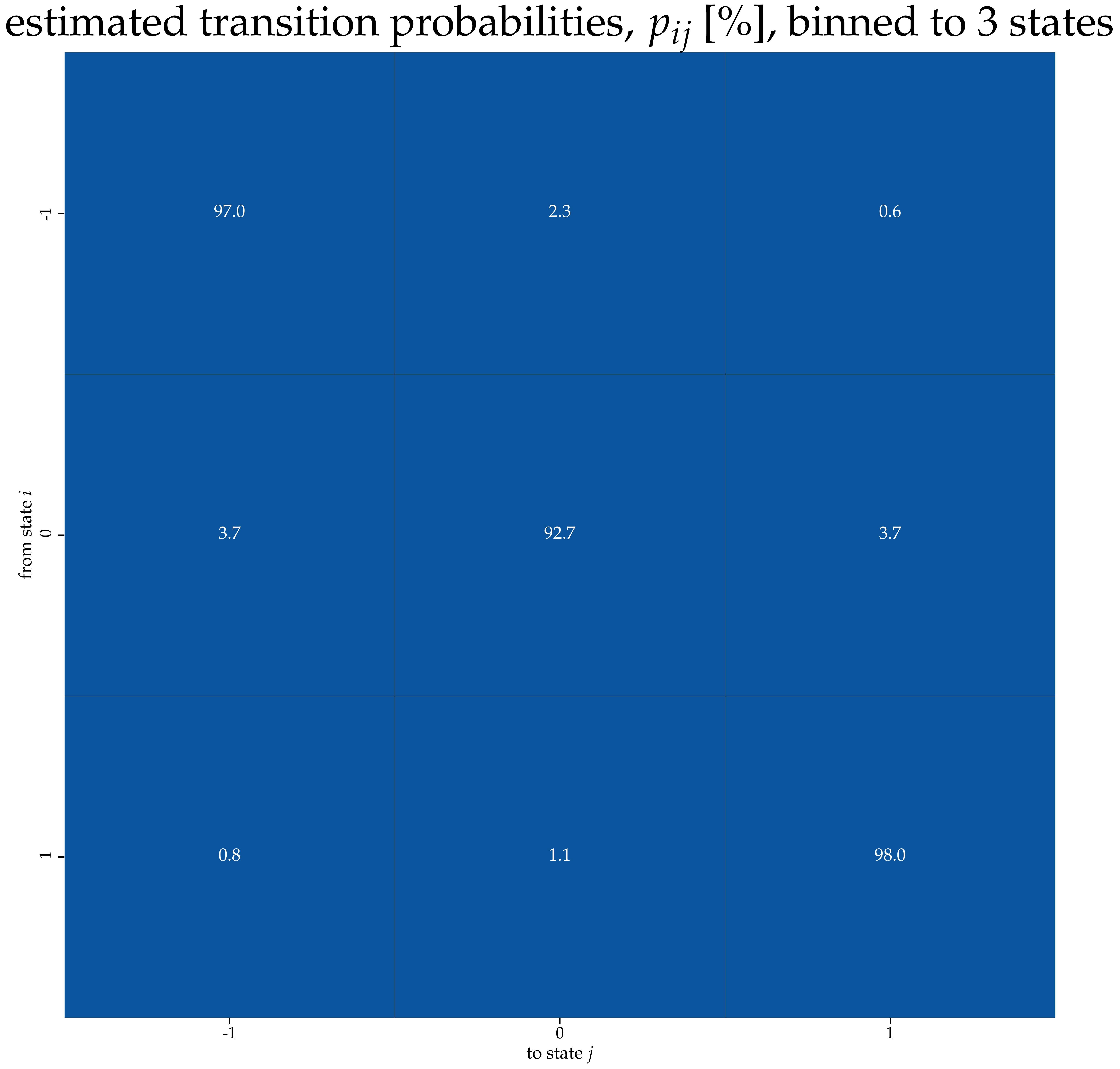}
\includegraphics[width=.49\linewidth]{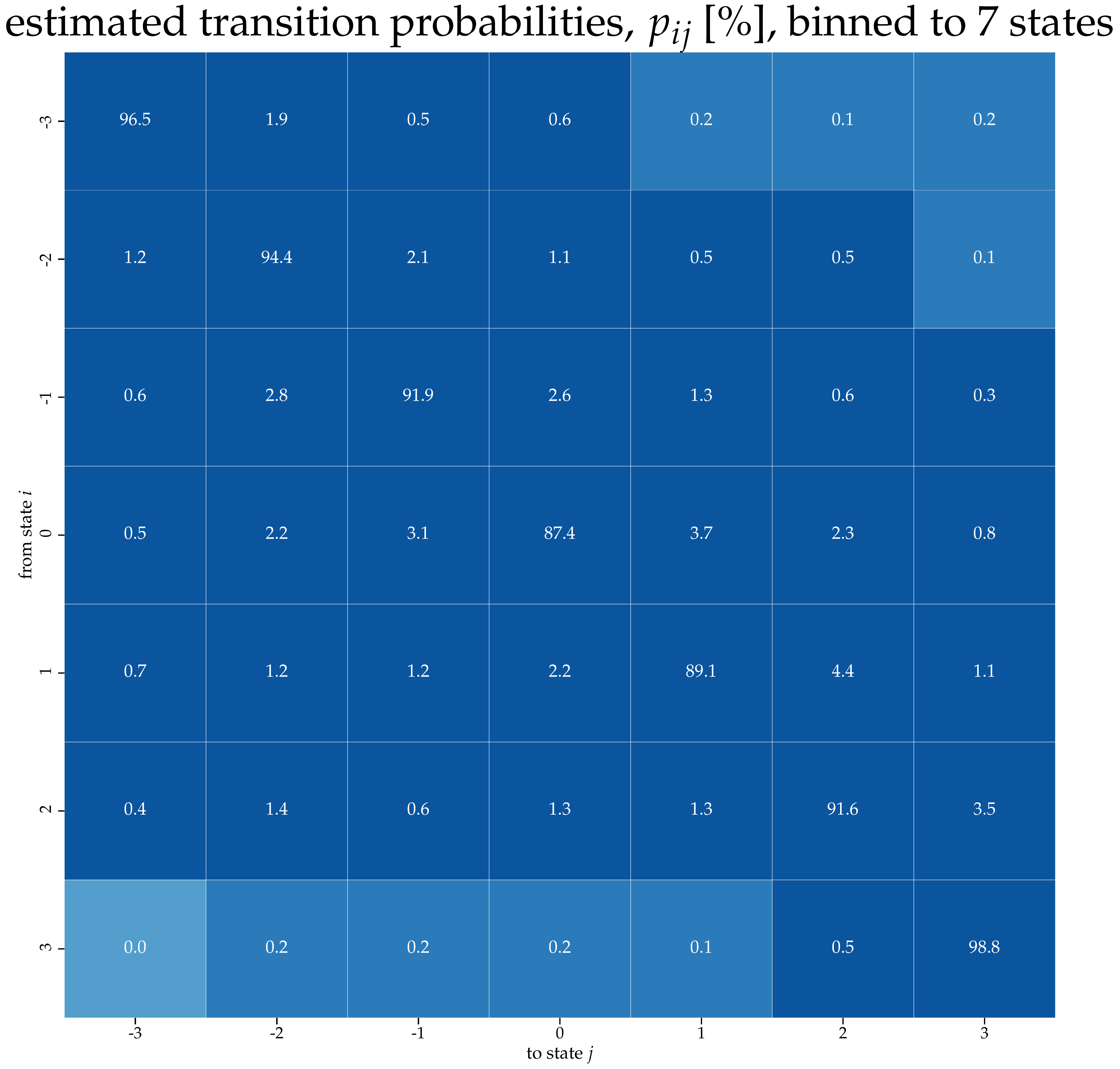}
\caption{The transition matrices derived from coarse-grained POLITY2 data reveal a similar structure to the original data. Transition probabilities are largest along the diagonal. We show matrices for coarse-graining to $3$ and $7$ states, respectively.}
\label{fig:transitionMatrix_binnings}
\end{figure}

In Fig.~\ref{fig:transitionMatrix_binnings}, we show the transition matrices estimated for the two coarse-grained data sets. We observe similar characteristics as for the original data: Transitions that do not change the state are most likely (i.e., diagonal elements of the matrices). Transitions that change the scores more tend to be less likely than transitions that change the score less. In both cases we also observe that the transition that the most democratic regime stays in this state ($s_3=+1$ and $s_7=+3$, respectively) are the most likely transitions. Overall, we observe ---in agreement with our discussion in the main manuscript--- that more extreme regimes tend to be more stable than mixed regimes.

\begin{figure}[b!]
\centering
\includegraphics[width=.49\linewidth]{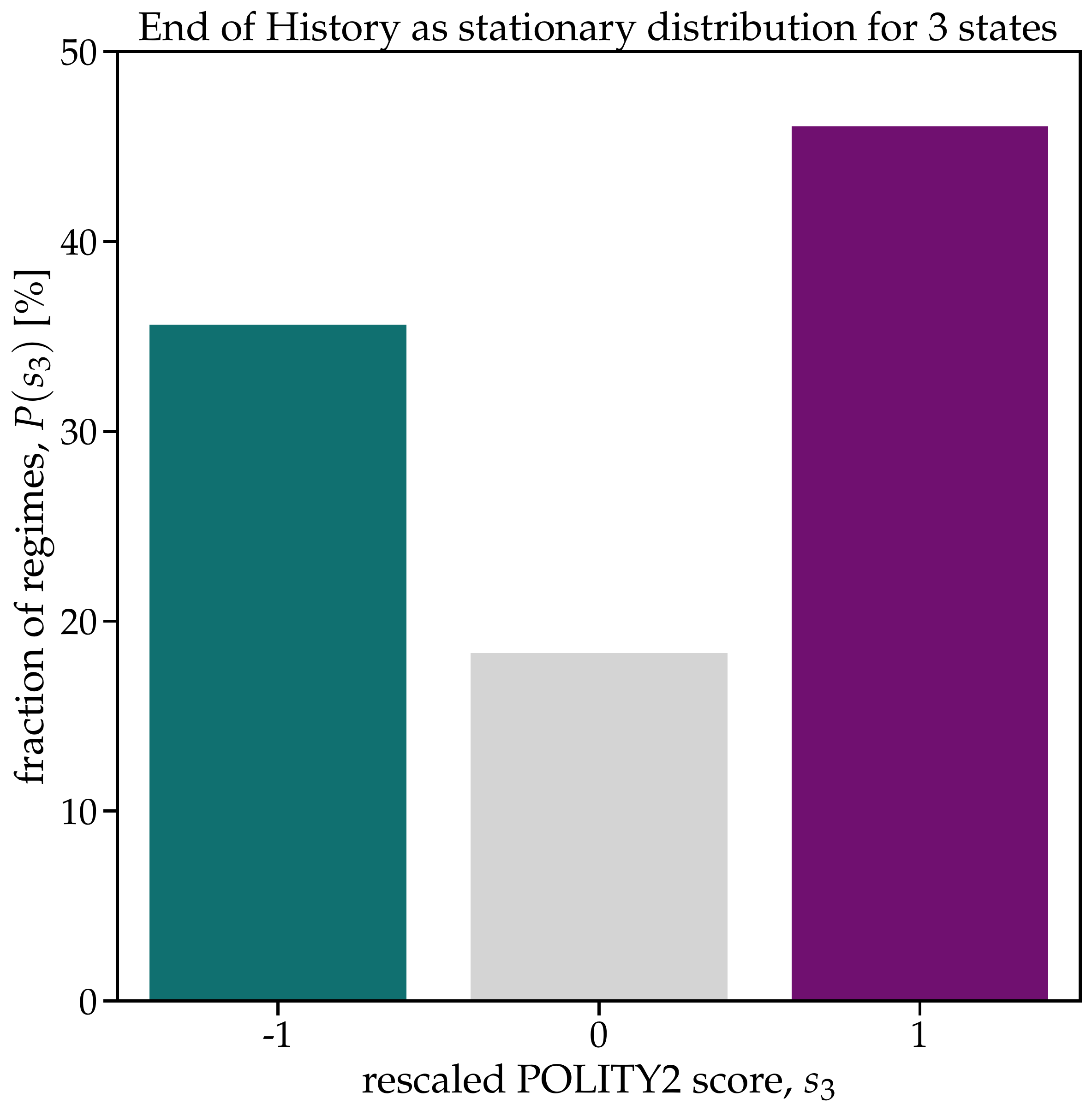}
\includegraphics[width=.49\linewidth]{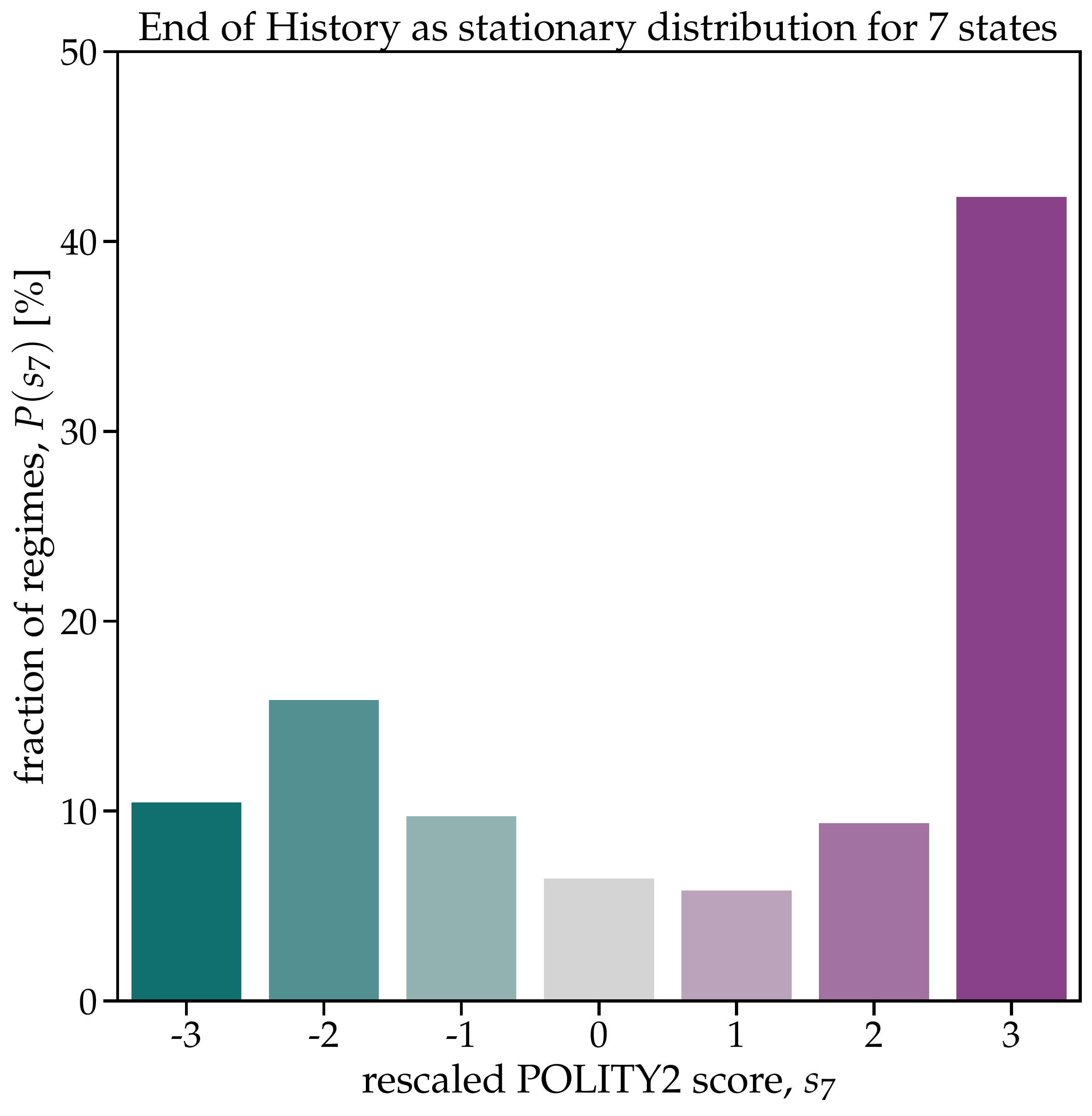}

\caption{The EoH for the coarse-grained POLITY2 data support the findings for the original data. A plurality but no majority of states are in the state representing the most democratic regimes ($s_3=+1$ and $s_7=+3$, respectively}. 
\label{fig:EoH_binnings}
\end{figure}

In Fig.~\ref{fig:EoH_binnings}, we show the steady state distributions that represent the EoH for both coarse-grained transition matrices. These distributions confirm our results from the main manuscript. In particular, we find that a plurality but no majority of all regimes is predicted to be in the most democratic state ($s_3=+1$ and $s_7=+3$, respectively). In both variants, we find a sizeable fraction of regimes to be autocratic.

\clearpage

\section{`End of history' for the V-Dem data}

In the main manuscript, we computed regime transition matrices from the POLITY2 data and used it to obtain an estimate for the EoH. In this section, we repeat this analysis with an alternative data source, which also describes regime transitions. The \emph{Varieties of Democracy} (V-Dem) is a measure of  democratic transitions~\cite{coppedge2019methodology} and is available under \url{https://www.v-dem.net/}.

We investigate the {\sc Electoral democracy index}, which is a measure of to what extent is the ideal of electoral democracy in its fullest sense achieved. In the V-Dem conceptual scheme, electoral democracy is understood as an essential element of any other conception of representative democracy (liberal, participatory, deliberative, egalitarian, or some other). The score ranges from $0.006$ to $0.926$ and we bin it into 21 states, to compare the EoH for the V-Dem score with the EoH for the POLITY2 score.

\begin{figure}[h]
\centering
\includegraphics[width=.5\linewidth]{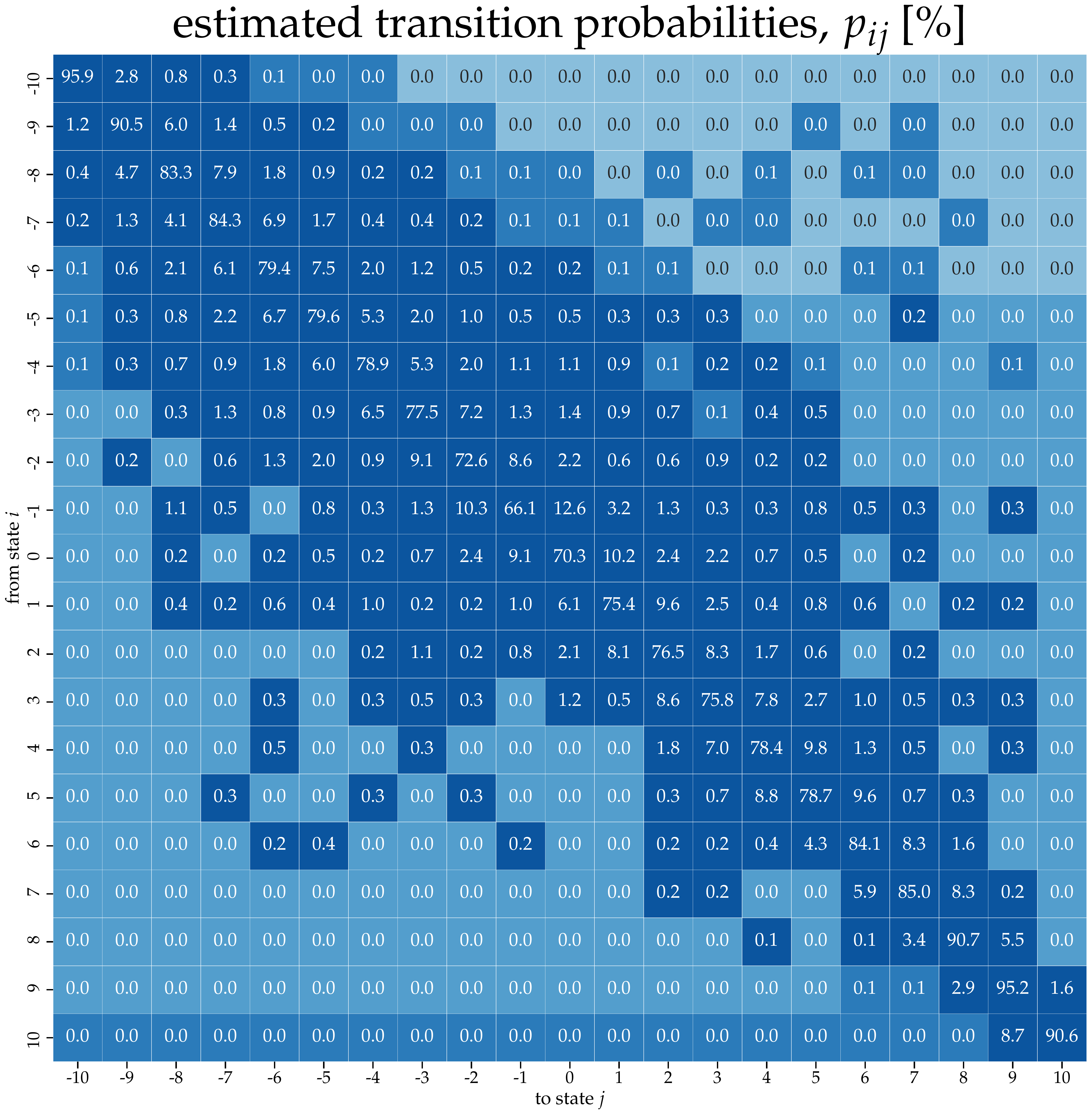}
\caption{The estimated transition matrix for the V-Dem data is similar to the one obtained from the POLITY2 data.}. 
\label{fig:transitionMatrix_VDem}
\end{figure}

In Fig.~\ref{fig:transitionMatrix_VDem}, we show the transition matrix obtained from the V-Dem data. We find similar features as in the POLITY2 data. In particular, regime transitions that do not change the state seem to be more likely than transitions that change the state strongly, leading to a drop-off of probabilities perpendicular to the diagonal. The transitions that has the highest probability is $-10$ to $-10$, indicating that full autocracies seem to be rather stable. The transition from $-1$ to $0$ is with $12.6\,\%$ probability the most likely state-changing transition.

\begin{figure}[t]
\centering
\includegraphics[width=.5\linewidth]{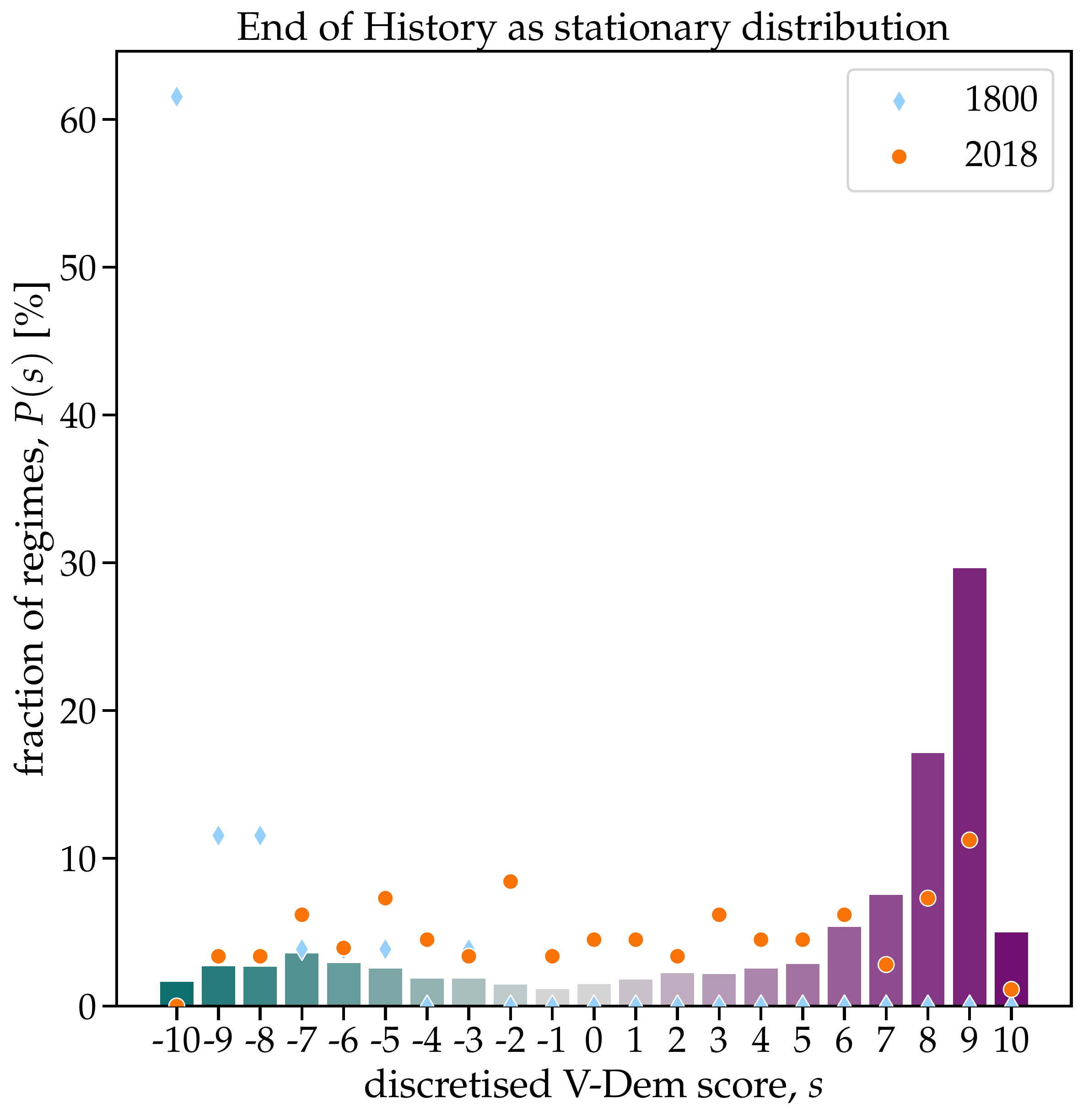}
\caption{The EoH for the V-Dem data support the findings for the POLITY2 data. }. 
\label{fig:EoH_VDem}
\end{figure}

In Fig.~\ref{fig:EoH_VDem}, we show the EoH as estimated from the V-Dem transition matrix. Similar to the POLITY2 data, we find that a plurality of countries is expected to have high V-Dem scores but a sizeable minority to have low V-Dem scores. In contrast to the POLITY2 score, we identify the second-largest category of democracies to have the highest probability.

\clearpage

\section{Temporal robustness of the `end of history'}

 In the main manuscript, we infer the Markov chain transition matrix $P$ from time series covering the POLITY2 score $s$ from 1800--2018. Using this transition matrix $P$, we then compute the EoH as the stationary distribution $\vec{\pi}$ associated with this transition matrix. This approach has the implicit assumption that there is one time-invariant transition matrix $P$, which is representative of the regime transition across the whole time period. 
 
  \begin{figure}[t]
\centering
\includegraphics[width=.45\linewidth]{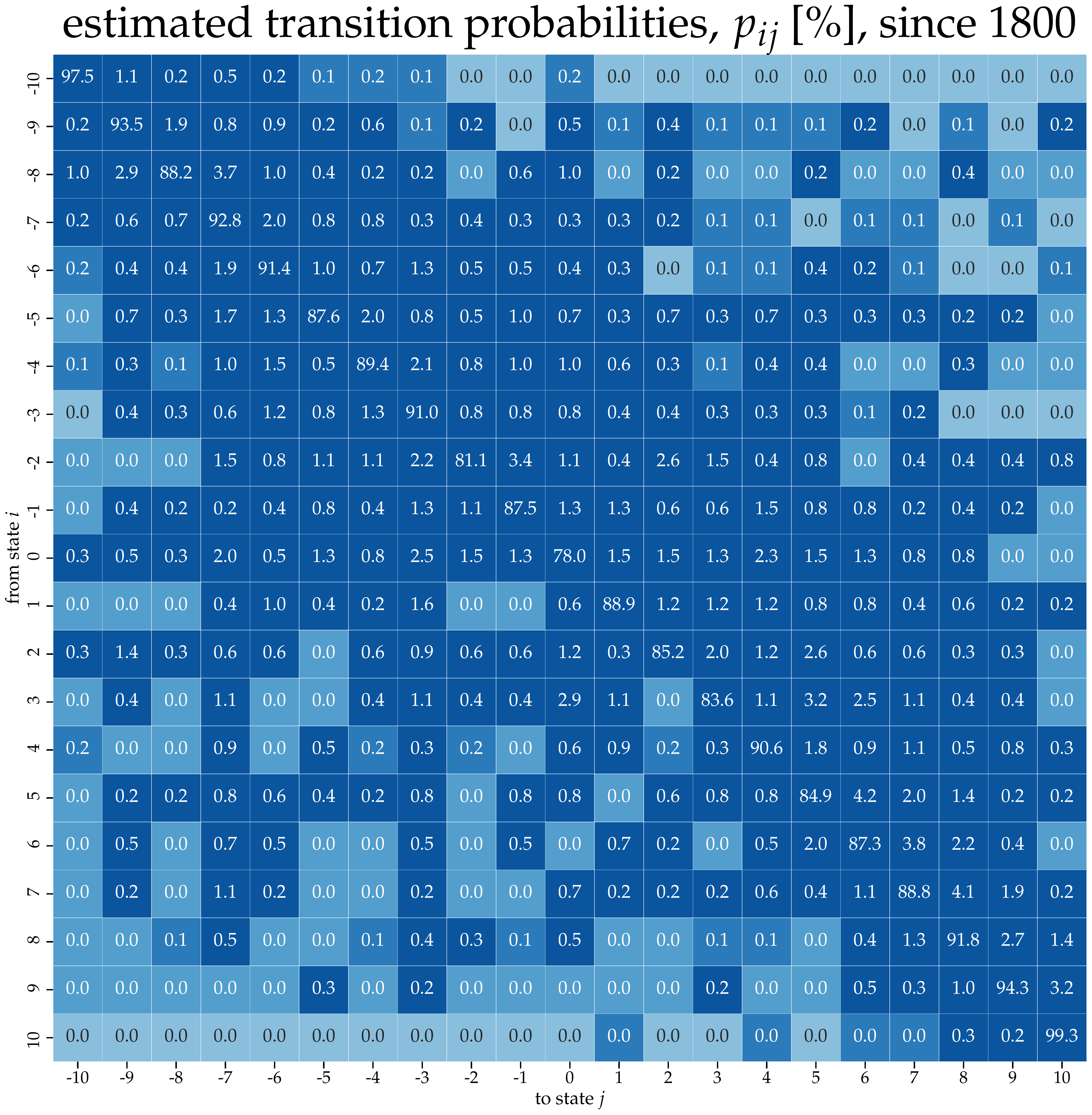}
\includegraphics[width=.45\linewidth]{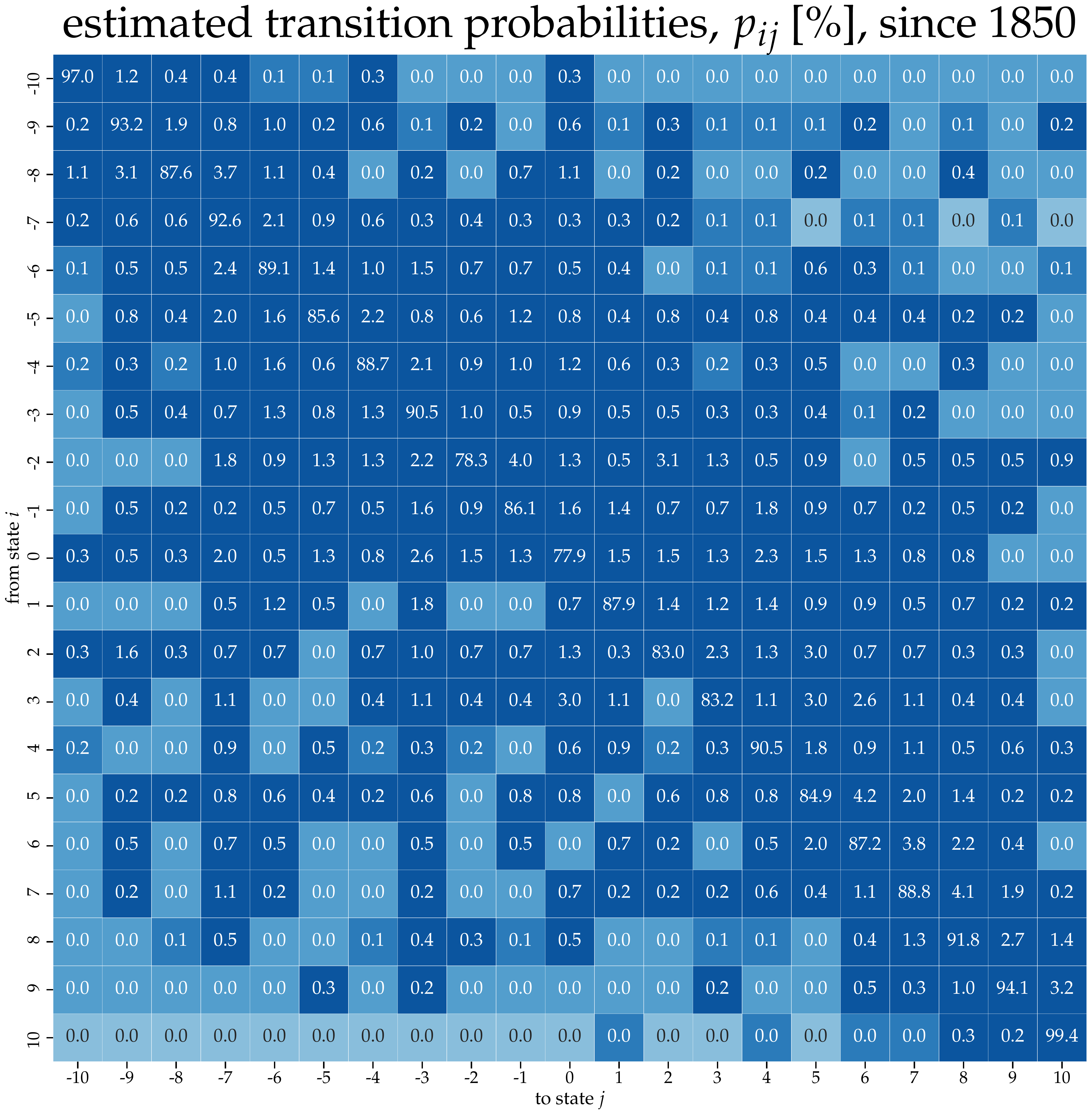}
\includegraphics[width=.45\linewidth]{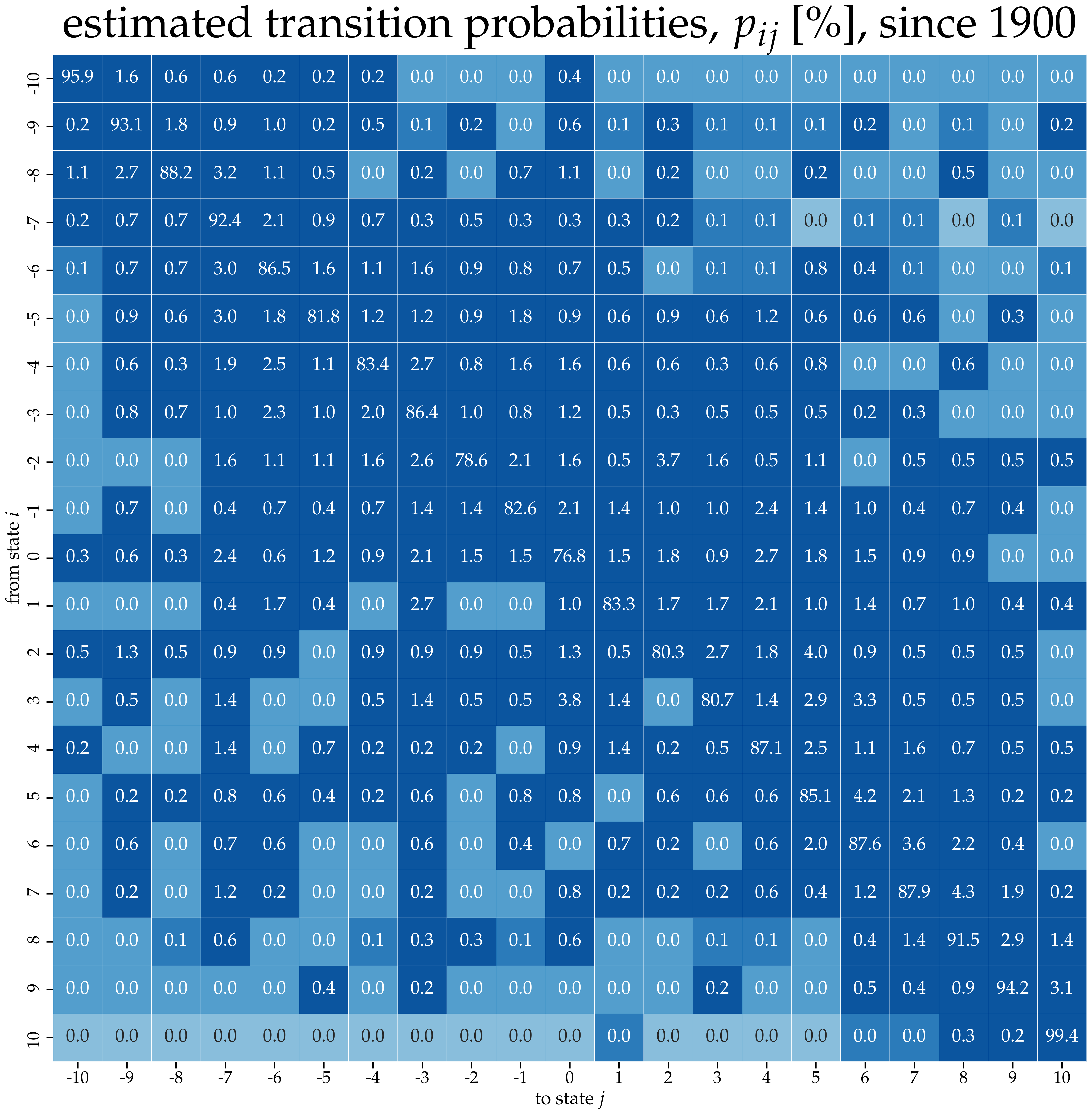}
\includegraphics[width=.45\linewidth]{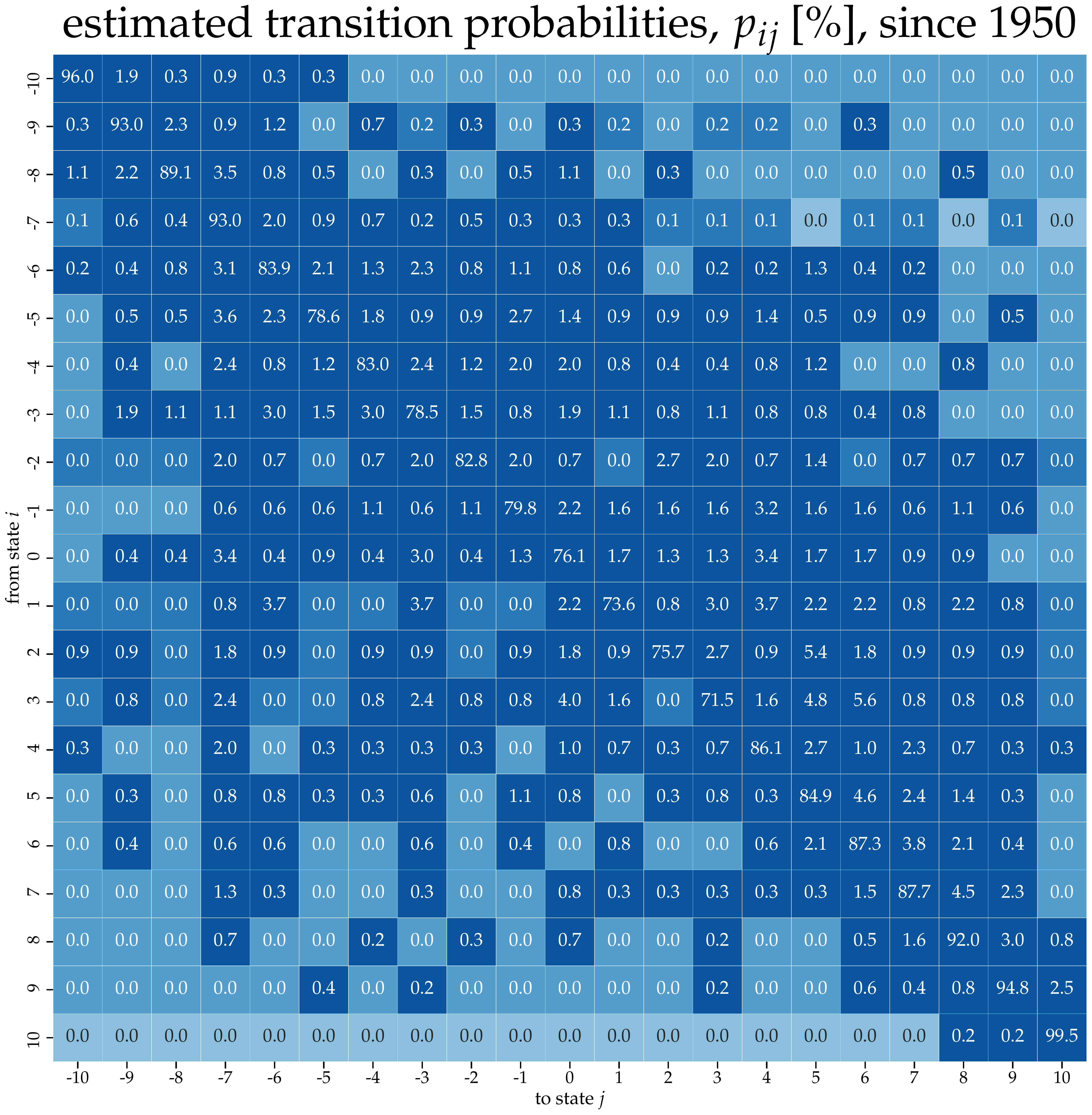}
\caption{The estimated transition matrix $P$ is largely invariant over time. We show the transition matrix for four time periods 1800--2018 (the whole data), 1850--2018, 1900--2018, and 1950--2018. The transition probabilities $p_{ij}$ are given in percent and colour-coded from low (bright) to high (dark).}
\label{fig:timeTransitionMatrices}
\end{figure}

 In this section, we will test to what extent choosing different time periods yields different transition matrices $P_t$, which in turn result in different EoHs. In particular, we study time periods 1800--2018 (the whole data), 1850--2018, 1900--2018, and 1950--2018. In Fig.~\ref{fig:timeTransitionMatrices}, we show the transition matrices estimated with the Bayesian approach, as outlined in the main manuscript. We find that the four matrices, while having slight differences, are largely similar, indicating that our assumption of time-invariance is, while being a simplifying assumption, is reasonable. 
 
\begin{figure}[t]
\centering
\includegraphics[width=.82\linewidth]{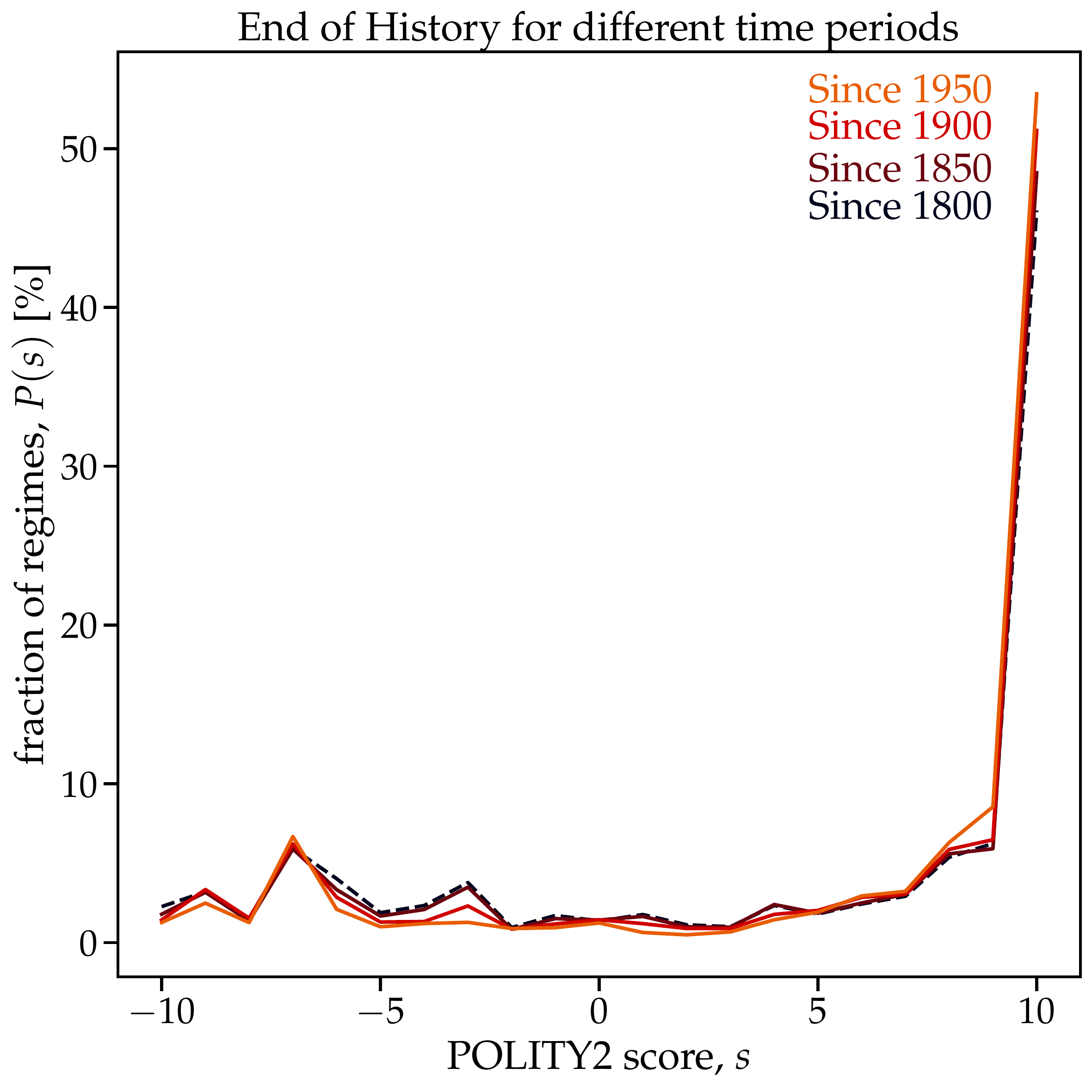}
\caption{The EoH is largely invariant for different time periods. 
We show the fraction of regimes that has has each POLITY2 score at the predicted EoH for four different time periods 1800--2018 (blue), 1850--2018 (purple), 1900--2018 (red), and 1950--2018 (orange). The fraction $P(s)$ of regimes in each POLITY2 score $s$ is very similar for each time period.}
\label{fig:time}
\end{figure}

 Next, we study the EoHs that are determined by the four different transition matrices. In Fig.~\ref{fig:time}, we show the EoH for each of the four time periods. We find that the estimated EoHs are very similar, indicating that the EoH is robust under these different choices for time periods. In particular, we find that for all four time periods the fraction of regimes as full democracies (i.e., $s=+10$) is approximately $50\,\%$. 
 
\begin{figure}[t]
\centering
\includegraphics[width=.82\linewidth]{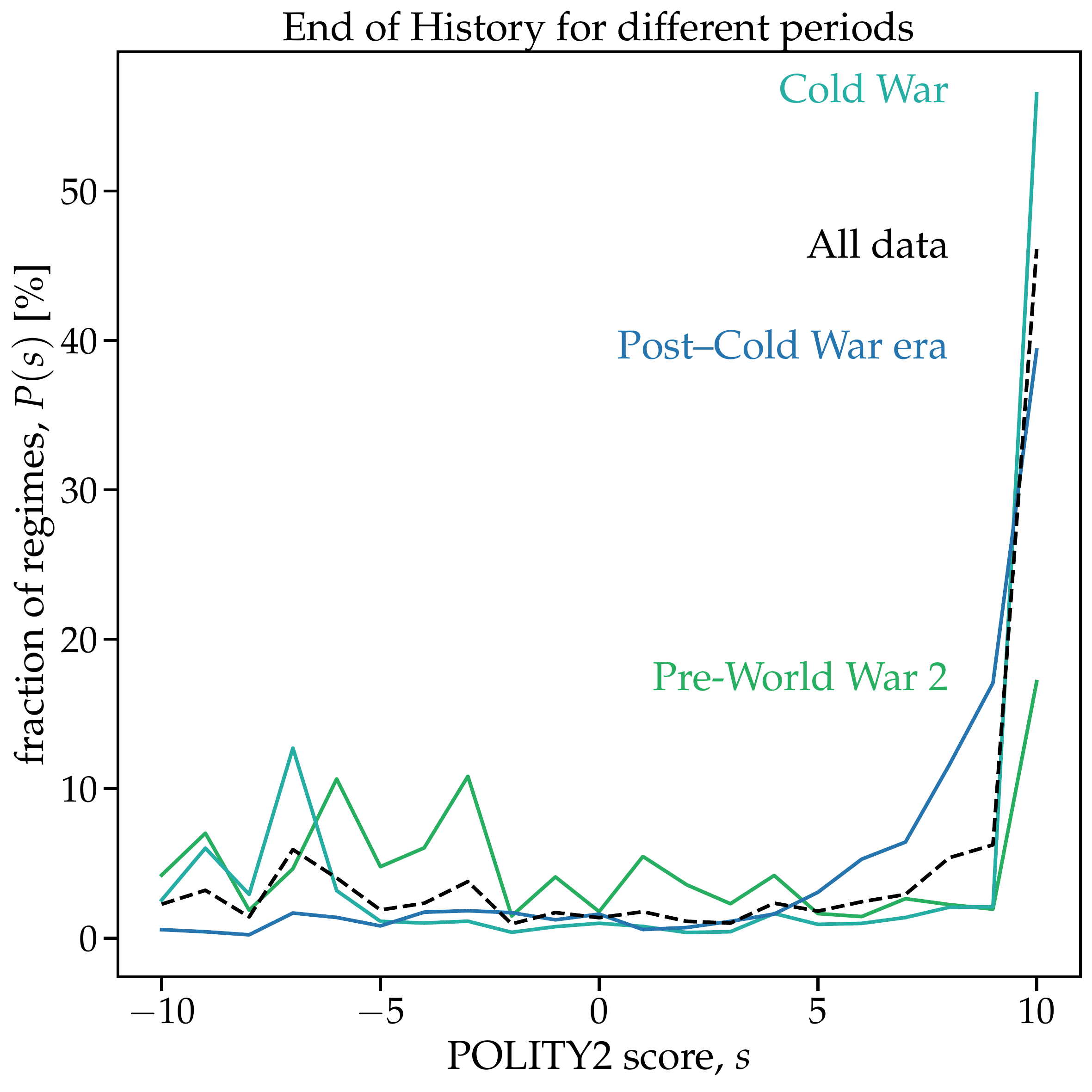}
\caption{The EoH is largely invariant for different time periods. 
We show the fraction of regimes that has has each POLITY2 score at the predicted EoH for four different time periods 1800--1939 (pre-World War 2; in green), 1945--1989 (Cold War; in cyan), and 1990--2018 (post-Cold War; in blue). For comparison, we also show the EoH for 1800--2018 (all data; in black). The fraction $P(s)$ of regimes in each POLITY2 score $s$ differs slightly for each time period but for all periods, we find that full democracies ($s=+10$) are most frequent.}
\label{fig:time2}
\end{figure}

 As an additional analysis, we also compute the EoH for three non-overlapping time periods (pre-World War 2, Cold War era, and post-Cold War era). We show the results in Fig.~\ref{fig:time2} and find that in all cases full democracies are predicted to be most frequent. We do, however, also observe some differences in the estimated EoHs. In particular, the frequency of regimes with negative POLITY2 scores is larger in the pre-World War 2 era, than in the post-Cold War era.

\clearpage

\section{`End of history' for frequentist approach}

In the main manuscript, we infer the Markov chain transition matrix $P$ with a Bayesian approach and use this to estimate the EoH. In this section, we compare this with the transition matrix and the associated EoH for the frequentist approach.

In a frequentist approach the maximum-likelihood estimation~\cite{grimmett2020probability} of the transition probability from state $i$ to state $j$ is then

\begin{align}
	\hat{p}^{\mathrm{F}}_{ij} = \frac{n_{ij}}{n_{i+}} = \frac{n_{ij}}{\sum_{j=1}^{N} n_{ij}}\,,
\label{eq:estimatorFreq}
\end{align}
where $n_{i+}$ is the total number of observed transitions starting in state $i$. 

The Bayesian approach is

\begin{align}
	\hat{p}_{ij}^{\mathrm{B}} = \frac{n_{ij} +\alpha_{ij}}{\sum_{j=1}^{N} (n_{ij}+\alpha_{ij})}\,,
\label{eq:estimatorBay}
\end{align}
where $\alpha_{ij}=1/N=1/21$ is the Dirichlet prior for which all transitions are equally likely.

\begin{figure}[b]
\centering
\includegraphics[width=.45\linewidth]{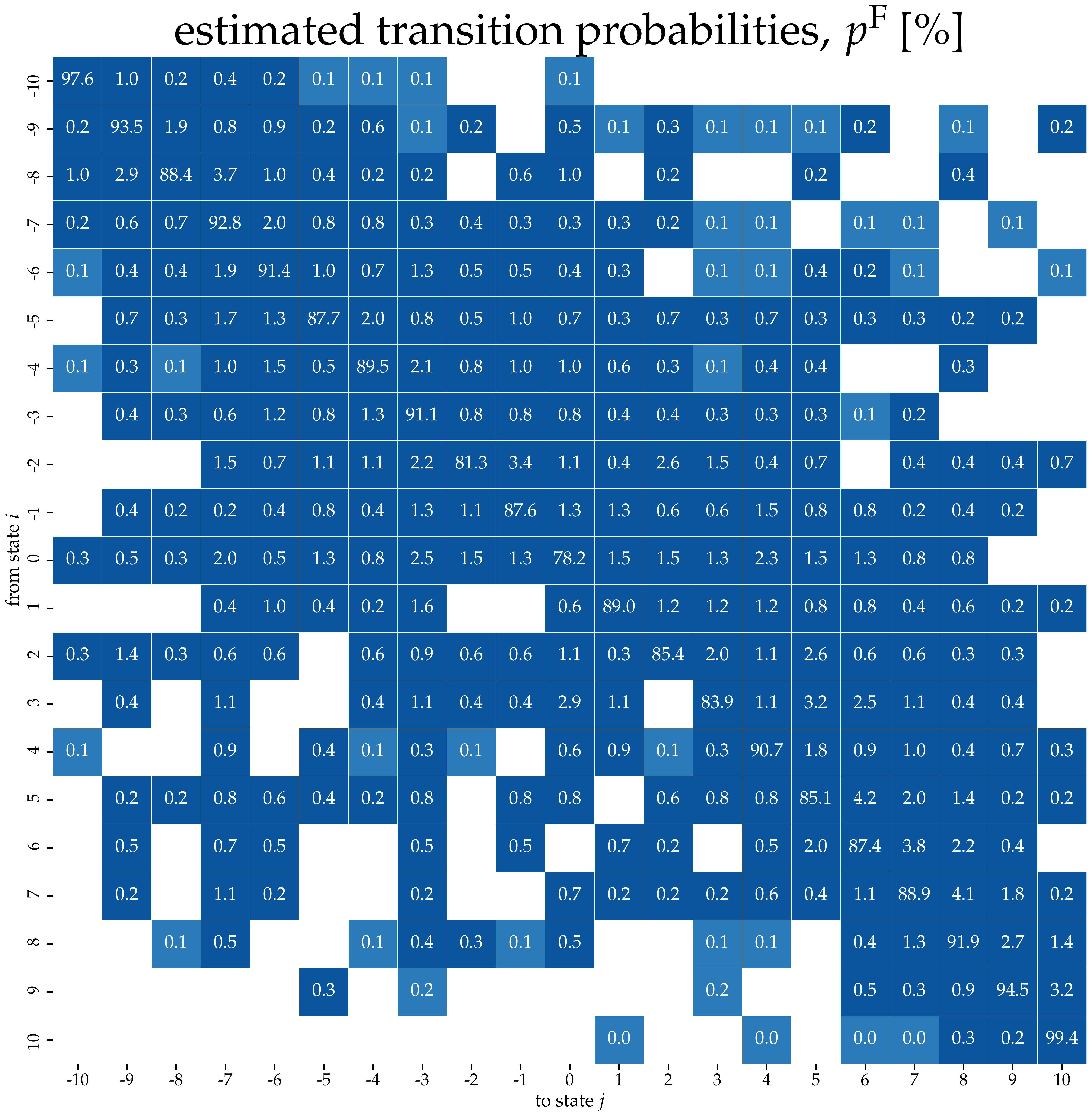}
\includegraphics[width=.45\linewidth]{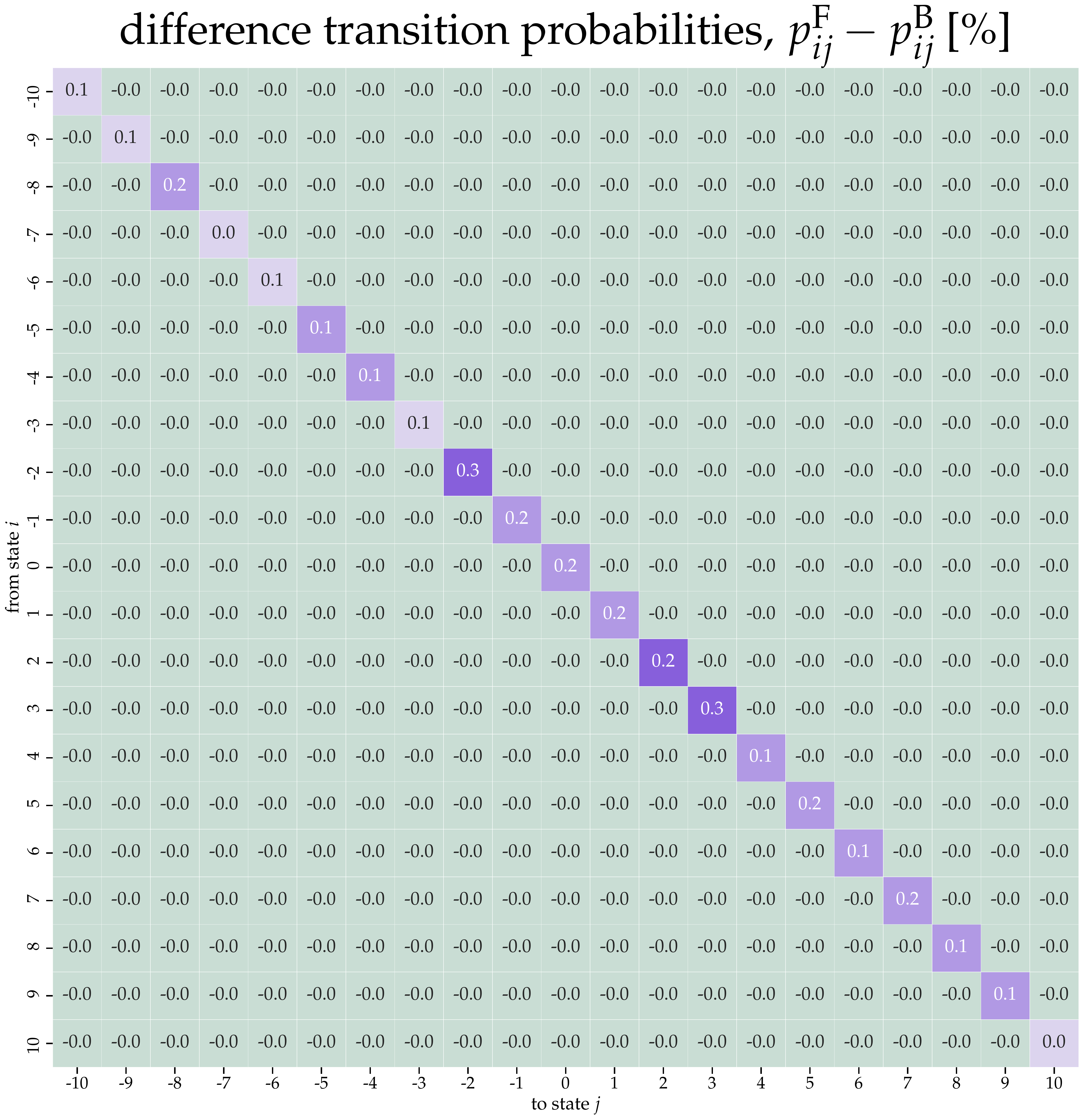}

\caption{The transition matrices inferred with the frequentist approach is similar to the one derived with the Bayesian approach. The main differences are that (1) never observer transitions have a zero probability and (2) that frequently observed transitions along the diagonal are more likely.}
\label{fig:frequentistTransitionMatrix}
\end{figure}

In Fig.~\ref{fig:frequentistTransitionMatrix}, we show the transition matrix estimated with a frequentist approach. We observe that transitions are most likely close to the diagonal, which represents transitions that change the POLITY2 score $s$ only slightly. Furthermore, we observe that transitions that have not been observed in our data (e.g., $-10$ to $-1$) have a transition probability of zero. In the Bayesian approach have a finite transition probability due to the prior. Next, we investigate the difference in the transition probabilities $p^{\mathrm{F}}_{ij} - p^{\mathrm{B}}_{ij}$ between these two approaches. We find that the off-diagonal elements are slightly larger in the Bayesian approach (as indicated by the green colour, note that the differences are less than $0.1\,\%$). The diagonal elements, in contrast, are slightly larger in the frequentist approach.

\begin{figure}[t]
\centering
\includegraphics[width=.43\linewidth]{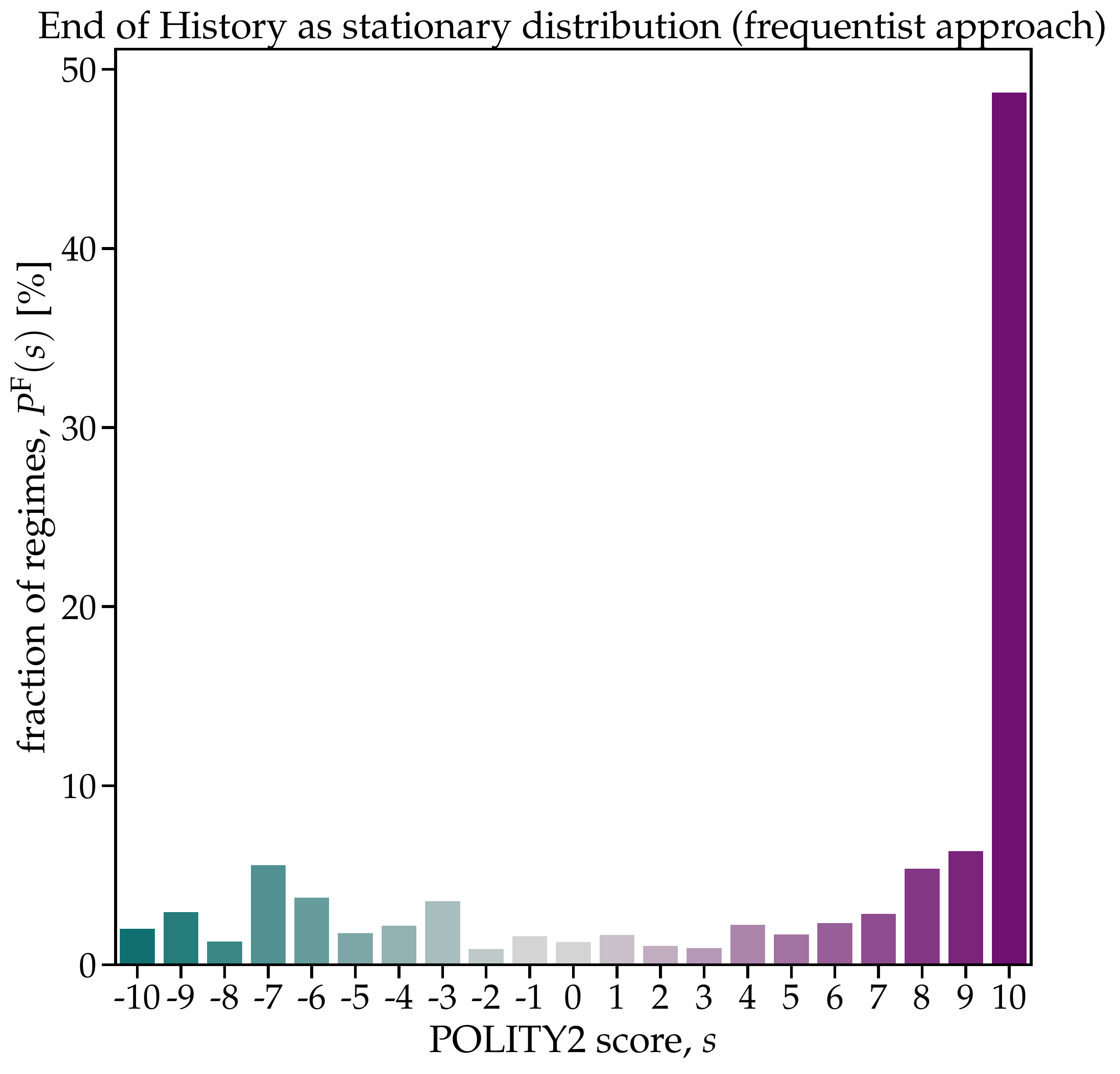}
\includegraphics[width=.48\linewidth]{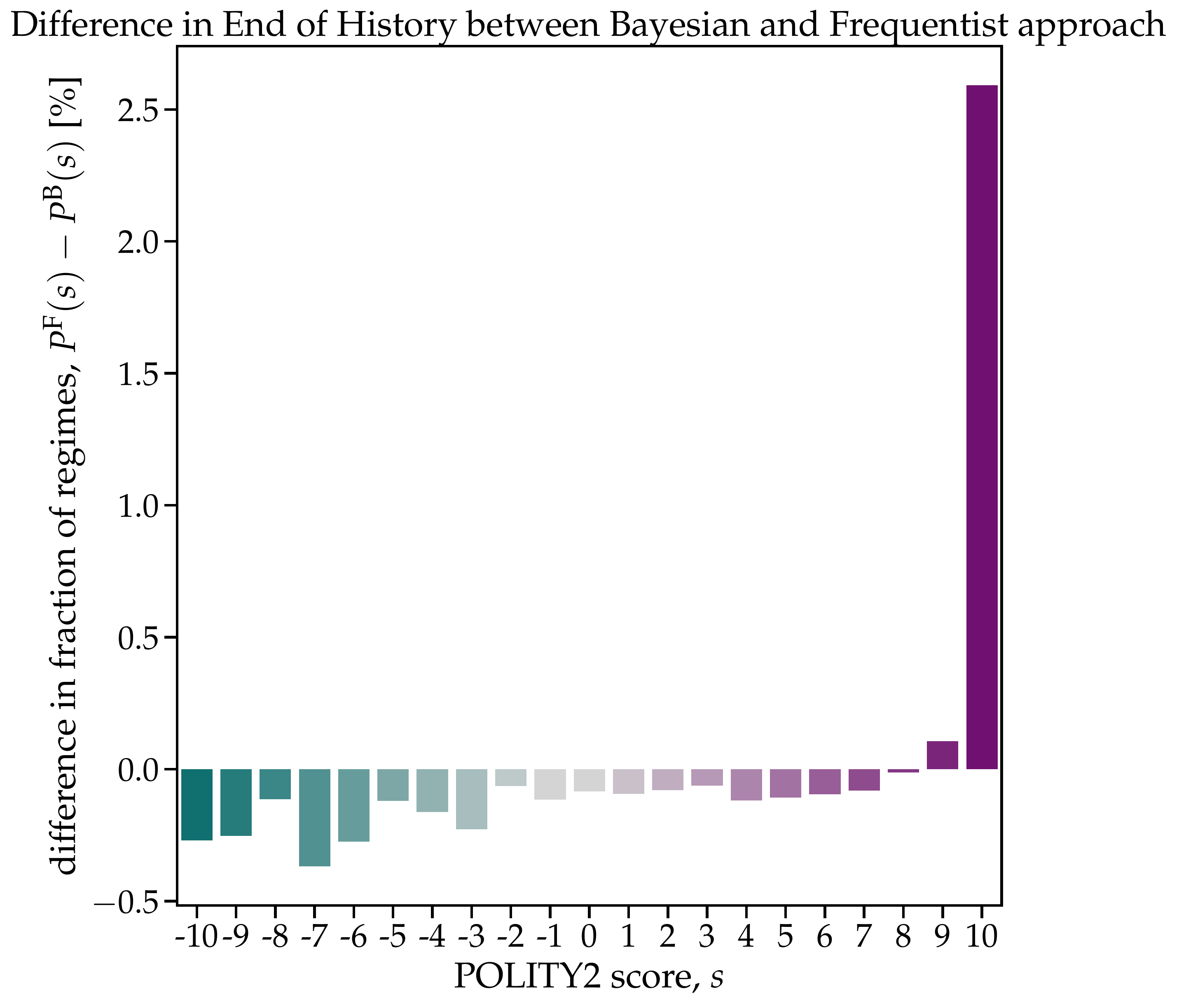}

\caption{The end of history estimated with the frequentist approach is similar to the one estimated with the Bayesian approach. The biggest difference is observed for the full democracies ($s=+10$), for which the frequentist approach results in an increase of  approximately $2.5\,\%$.}
\label{fig:frequentistEndOfHistory}
\end{figure}

In Fig.~\ref{fig:frequentistEndOfHistory}, we show the EoH for the frequentist approach and compare it the Bayesian approach. We find that our main findings hold. In particular, a plurality of regimes are expected to be full democracies (i.e., $s=+10$). In fact, we find that the frequentist approach results in an increase by $2.5\,\%$ for this state. The state $s=+9$ is also slightly increased, whereas all other states are reduced.

Overall, these findings indicate that the Bayesian approach makes rarely observed transitions more likely, which results in an EoH at which autocracies are slightly more abundant. We note that the Bayesian approach is fruitful ---despite the very similar result to the frequentist approach--- because it guarantees that the obtained Markov chain transition matrix yields a unique steady-state distribution $\vec{\pi}$ (i.e., EoH), which is not the case for the frequentist approach.

\clearpage

\section{Survival probability from the empirical cumulative distribution function}

In the main manuscript, we estimate the survival function $S(t)$ of political regimes from simulated regime-transition time series. In particular, we simulate $r=10^5$ political regimes and obtain their life times as hitting times. We use the non-parametric Kaplan--Meier estimator~\cite{kaplan1958nonparametric}. 

It is also possible to illustrate these survival probabilities directly from the simulated regime-transition time series. For this, we compute the empirical cumulative distribution function (ECDF)

\begin{align}
\hat{F}_n(t) = \frac{1}{n} \sum _{t=1}^r X_i \leq t\,,	
\end{align}
where $(X_1,X_2,\dots, X_r)$ are the life times (i.e., hitting times) of political regimes.

\begin{figure}[b]
\centering
\includegraphics[width=.9\linewidth]{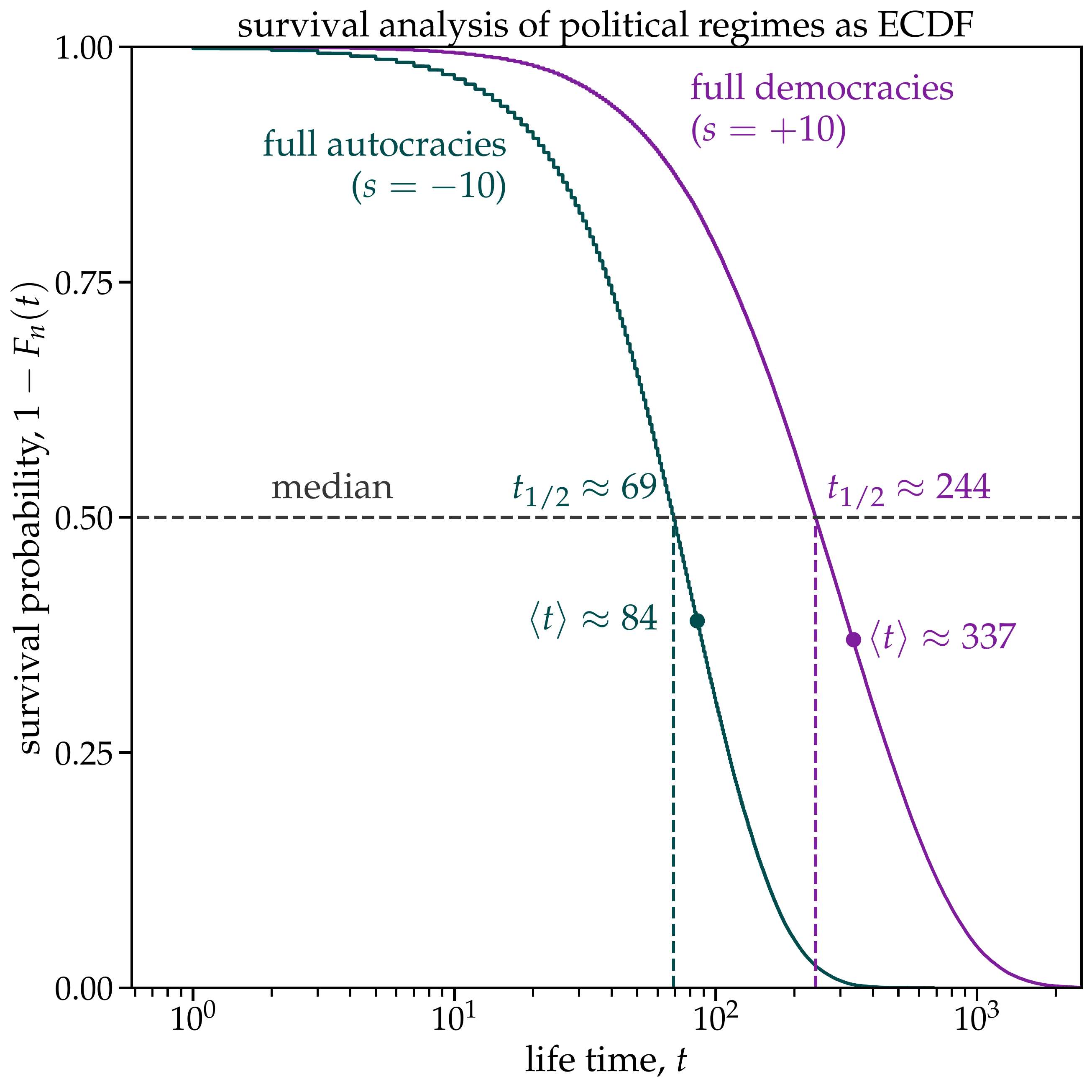}
\caption{The survival curves for the ECDF is almost identical to the ones obtained from the Kaplan--Meier estimator.}
\label{fig:surivalCurves_ECDF}
\end{figure}

In Fig.~\ref{fig:surivalCurves_ECDF}, we show the survival function as ECDF and find that it is virtually indistinguishable from the Kaplan--Meier estimates, as we expect from data without censoring.

\clearpage

\section{Best fit for the expected POLITY2 score change $\Delta s(s)$}

In the main manuscript, we identify that that expected POLITY2 score change $\Delta s$ as a function of the POLITY2 score $s$ follows roughly a polynomial of order $k=3$. Here, we wont to verify that such a cubical polynomial describes the data best.

\begin{figure}[t]
\centering
\includegraphics[width=.7\linewidth]{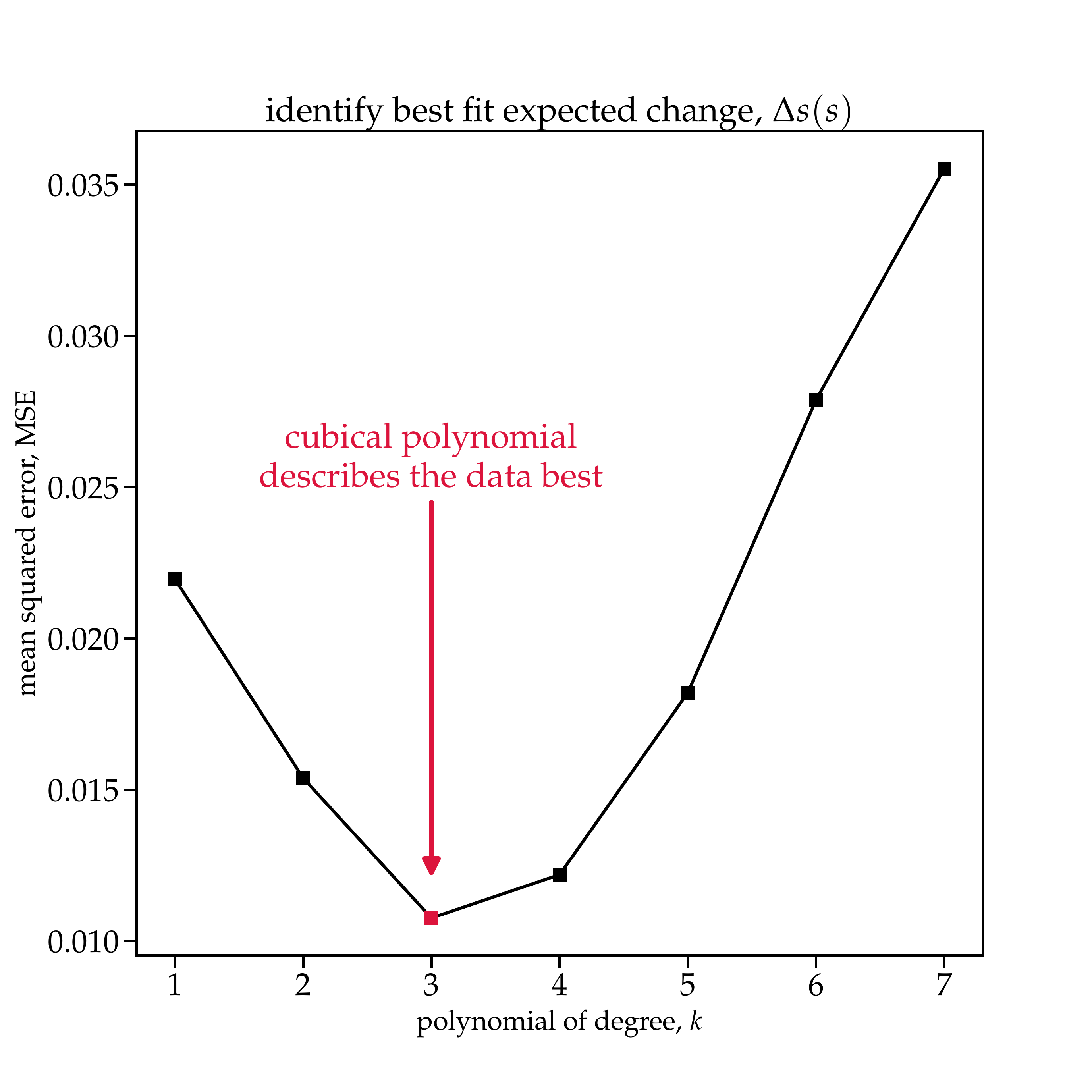}
\caption{The expected POLITY2 score change $\Delta s$ as a function of the POLITY2 score $s$ can be best described by a polynomial of order $k=3$.}
\label{fig:polynomialOrder}
\end{figure}

We perform a cross-validation to compare the fit of polynomials of different order to the data. In particular, we split the data into $80\,\%$ training data and $20\,\%$ test data such that we fit the polynomials to the training data and predict the test data. To compare the  performance of the different prediction we compute a mean squared error (MSE)

\begin{align}
	\text{MSE} = \frac{1}{n} \sum_{i=1}^n \left( \Delta s (s) - \hat{\Delta} s (s) \right)^2\,,
\end{align}  
where $\hat{\Delta} s (s) $ is the POLITY2 score change $\Delta s$ as predicted by a certain polynomial. In Fig.~\ref{fig:polynomialOrder}, we show the MSE for polynomials of order $k=1,\dots, 7$. We find the smallest MSE for $k=3$, which demonstrates that a cubical polynomial is an appropriate fit.

\clearpage

\section{Model selection for Markov chains}

In the main manuscript, we model the time-series data as a first-order Markov chain. In principle, it would also be possible to model the data with a higher-order Markov chain (i.e., the transition probabilities depend not only on the current state but also on further states). In this section, we demonstrate that a first-order Markov chain is appropriate choice.

We use a model-selection technique outlined in~\cite{petrovic2022learning} to learn the Markov order of the analysed time series. For this, we compute the Akaike information criterion

\begin{align}
	\text{AIC} = 2k - 2 \ln (\hat{L})\,,
\end{align}
and the Bayesian information criterion
\begin{align}
	\text{BIC} = 2\ln (n) - 2 \ln (\hat{L})\,,
\end{align}
where $\hat{L}$ is the maximised likelihood function, $n$ the number of data points, and $k$ the number of parameters of the model. In general, models with smaller values of the information criteria are preferred over models with larger values. 

\begin{figure}[t]
\centering
\includegraphics[width=.5\linewidth]{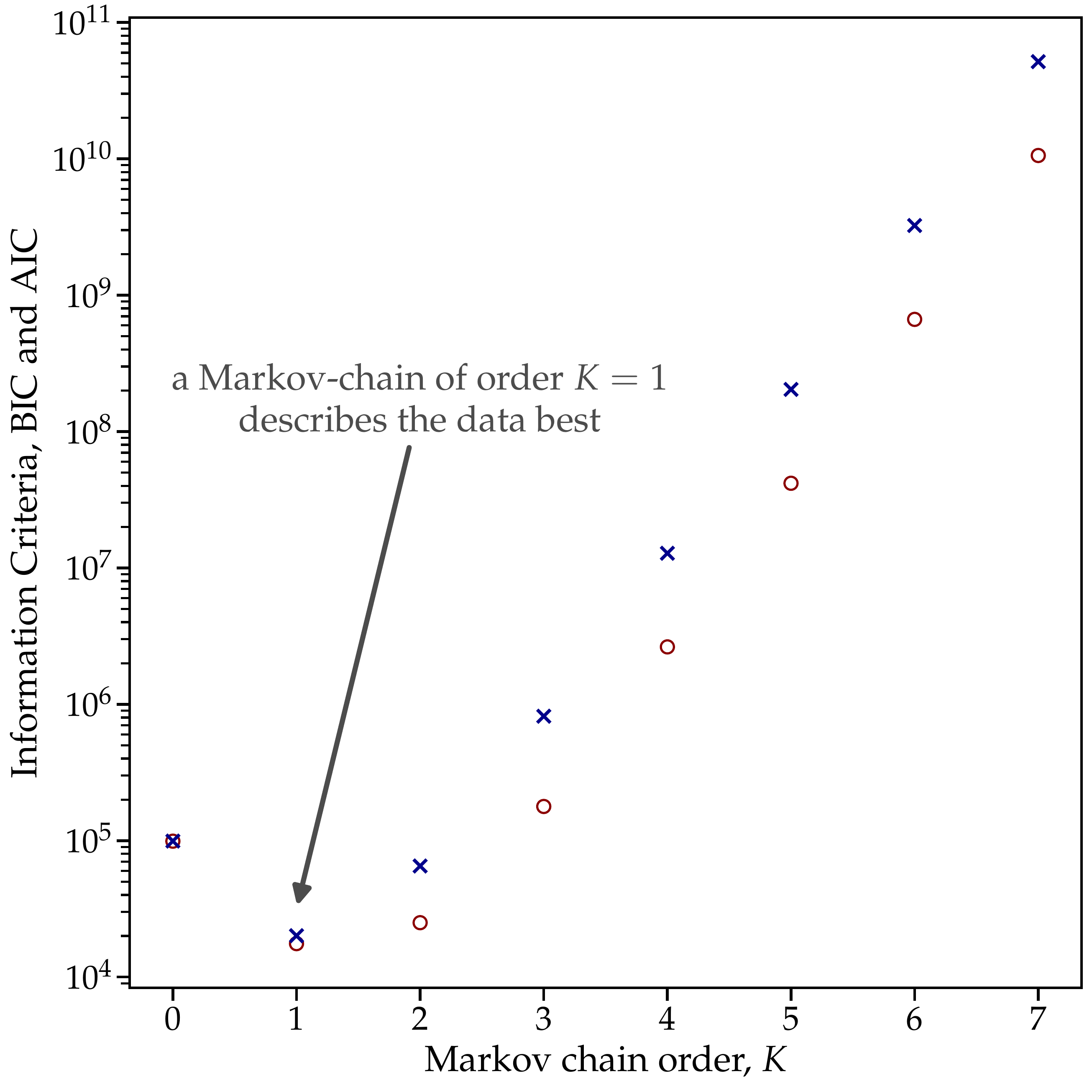}
\caption{The time series data is best described by a Markov chain of order $K=1$, which indicates that the memoryless assumption is verified. We show the AIC and BIC for Markov chain models of orders $K=0,\dots, 7$ and identify that both information criteria a smallest for a Markov chain of oder $K=1$.}
\label{fig:markovOrder}
\end{figure}

In Fig.~\ref{fig:markovOrder}, we show the AIC and BIC for Markov-chain models of orders $K=0,\dots, 7$. As both information criteria yield a minimum at order $K=1$, we can conclude that the memoryless assumption is a reasonable assumption.

We highlight that the authors in~\cite{petrovic2022learning} outline further (mathematically more advanced) tests for the appropriate Markov-order of data. All of these tests verify that a first-order Markov chain is most appropriate for the data. 

\clearpage

\section{Counterfactuals}

In the main manuscript, we compute the EoH for the truly observed time-series data. Our quantitative analysis allows us to test the robustness of our results under changes in the data. For example, we can test certain counterfactuals of the form `if a certain regime transition would (not) have happened, how would the predicted EoH be changed.'

In particular, we investigate two counterfactuals
\begin{itemize}
	\item no collapse of the USSR (i.e., we remove all post-soviet states from the data set)
	\item larger democratisation during the Arab Spring (i.e., we select twelve countries that were the centre of the Arab Spring and set their POLITY2 scores from 2011 onwards to $+10$)
\end{itemize}

For both cases, we re-compute the EoH and observe fairly moderat changes in comparison to the original result. For the USSR counterfactual, we find that the proportion of full democracies ($s=+10$) increases from $46\,\%$ to $47\,\%$. For the Arab Spring counterfactual, we identify that the proportion of full democracies increases from $46\,\%$ to $51\,\%$. Both cases indicate that, as expected, such major changes do influence the steady state distribution but the overall finding of full democracies being most frequent at the EoH is unchanged.  

\begin{figure}[t]
\centering
\includegraphics[width=.82\linewidth]{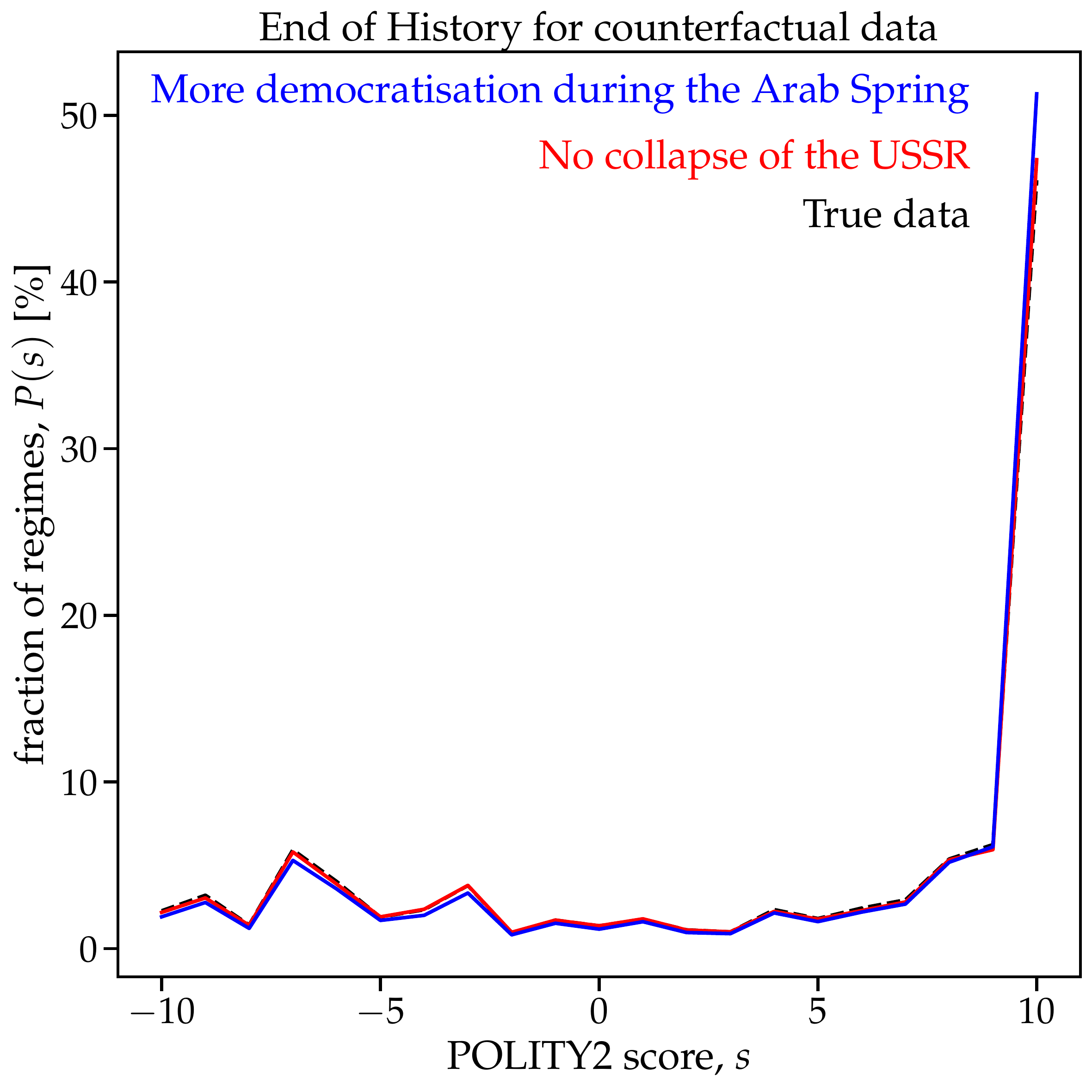}
\caption{The EoH is largely invariant under counterfactual perturbations of the data. 
We show the fraction of regimes that has has each POLITY2 score at the predicted EoH for two counterfactual perturbations of the data: no collapse of the USSR (red) and more democratisation during the Arab Spring (blue). For comparison, we also show the EoH for the true data (black). The fraction $P(s)$ of regimes in each POLITY2 score $s$ is slightly altered for the counterfactual perturbations but we find that full democracies ($s=+10$) are most frequent in all cases.}
\label{fig:counterfactuals}
\end{figure}

\clearpage

\bibliography{./democracyState}

\bibliographystyle{unsrt}